\documentclass[journal]{IEEEtran}

\usepackage[T1]{fontenc}
\usepackage{color}
\usepackage[english]{babel}
\usepackage{array}
\usepackage{prettyref}
\usepackage{booktabs}
\usepackage{url}
\usepackage{multirow}
\usepackage{amsthm}
\usepackage{amsmath}
\usepackage{amssymb}
\usepackage{graphicx}
\usepackage{flushend}
\usepackage{cite}
\ifCLASSOPTIONcompsoc
\usepackage[caption=false,font=normalsize,labelfont=sf,textfont=sf]{subfig}
\else
\usepackage[caption=false,font=footnotesize]{subfig}
\fi
\usepackage{cite}
\usepackage{MnSymbol}

\providecommand{\tabularnewline}{\\}

\theoremstyle{plain}
\newtheorem{thm}{\protect\theoremname}
\theoremstyle{plain}
\newtheorem{prop}[thm]{\protect\propositionname}

\providecommand{\propositionname}{Proposition}
\providecommand{\theoremname}{Theorem}

\ifCLASSINFOpdf
\else
\fi
\usepackage[normalem]{ulem}
\usepackage{hyperref}

\begin{document}

%
\title{Media Query Processing For The Internet-of-Things: Coupling Of Device Energy Consumption And\ Cloud Infrastructure Billing}

\author{Francesco Renna, Joseph Doyle, Vasileios Giotsas and Yiannis Andreopoulos~\IEEEmembership{Senior Member,~IEEE}
\thanks{F. Renna is with the Department of Applied Mathematics and Theoretical Physics, University of Cambridge, Wilberforce Road, Cambridge, CB3 0WA, UK (e-mail: fr330@cam.ac.uk)}
\thanks{J. Doyle  and Y. Andreopoulos are with the Electronic and Electrical Engineering Department, University College London, Roberts Building, Torrington Place, London, WC1E 7JE, UK (e-mail:
\{j.doyle,  i.andreopoulos\}@ucl.ac.uk)}
\thanks{V. Giotsas is with Dithen Ltd., \href{www.dithen.com\#www.dithen.com}{www.dithen.com}, 843 Finchley Road, London NW11 8NA, UK (e-mail: v.giotsas@dithen.co.uk)}
\thanks{This work was supported in part by the European Union (Horizon 2020 research and innovation programme under the Marie Sk{\l}odowska-Curie grant agreement No 655282 -- F. Renna), EPSRC (grants EP/M00113X/1 and EP/K033166/1) and Innovate UK (project ACAME, grant 131983).}}


%


\maketitle

\begin{abstract}
Audio/visual recognition and retrieval applications have recently garnered significant attention within Internet-of-Things (IoT) oriented services, given that video cameras and audio processing chipsets are now ubiquitous even in low-end embedded systems. In the most typical scenario for such services, each device extracts audio/visual features and compacts them into feature descriptors, which comprise media queries. These queries are uploaded to a remote cloud computing service that performs content matching for classification or retrieval applications. Two of the most crucial aspects for such services are: \textit{(i)} controlling the device energy consumption when using the service; \textit{(ii)} reducing the billing cost incurred from the cloud infrastructure provider.  In this paper we derive analytic conditions for the optimal coupling between the device energy consumption and the incurred cloud infrastructure billing. Our framework encapsulates: the energy consumption to produce and transmit audio/visual queries, the billing rates of the cloud infrastructure, the number of devices concurrently connected to the same cloud server, {the query volume constraint of each cluster of devices,} and the statistics of the query data production volume per device. Our analytic results are validated via a deployment with: \textit{(i)} the device side comprising compact image descriptors (queries) computed on Beaglebone Linux embedded platforms and transmitted to Amazon Web Services (AWS) Simple Storage Service; \textit{(ii)} the cloud side carrying out image similarity detection via \ AWS\ Elastic Compute Cloud (EC2) instances, with the AWS Auto Scaling being used to control the number of instances according to the demand. 
\end{abstract}

\begin{IEEEkeywords}
visual search, internet-of-things, cloud computing, analytic
modeling
\end{IEEEkeywords}

%
\IEEEpeerreviewmaketitle

\section{Introduction}
Most of the envisaged applications and services
for wearable sensors, smartphones, tablets or portable computers in
the next ten years will involve analysis of audio/visual streams for event,
action, object or user recognition, recommendation services and context awareness, etc. \cite{siewiorek2012generation,5754008,6616112,6616113,leung2013cloud,soyata2012cloud,girod2011mobile,serra2010audio}.
Examples of early commercial services in this domain include Google
Goggles, Google Glass object recognition, Facebook automatic face
tagging \cite{becker2008evaluation}, Microsoft's Photo Gallery face
recognition, as well as technology described in recent publications
from Google, Siemens and others
\footnote{See ``A Google Glass app knows what you're looking at'' MIT Tech.
Review (Sept. 30, 2013) and EU projects SecurePhone \cite{sellahewa2005wavelet,bredin2006detecting}
and MoBio \cite{poh2010evaluation,Marcel_CVPR_2010}.}. 

\begin{figure}
\begin{centering}
\includegraphics[scale=0.15]{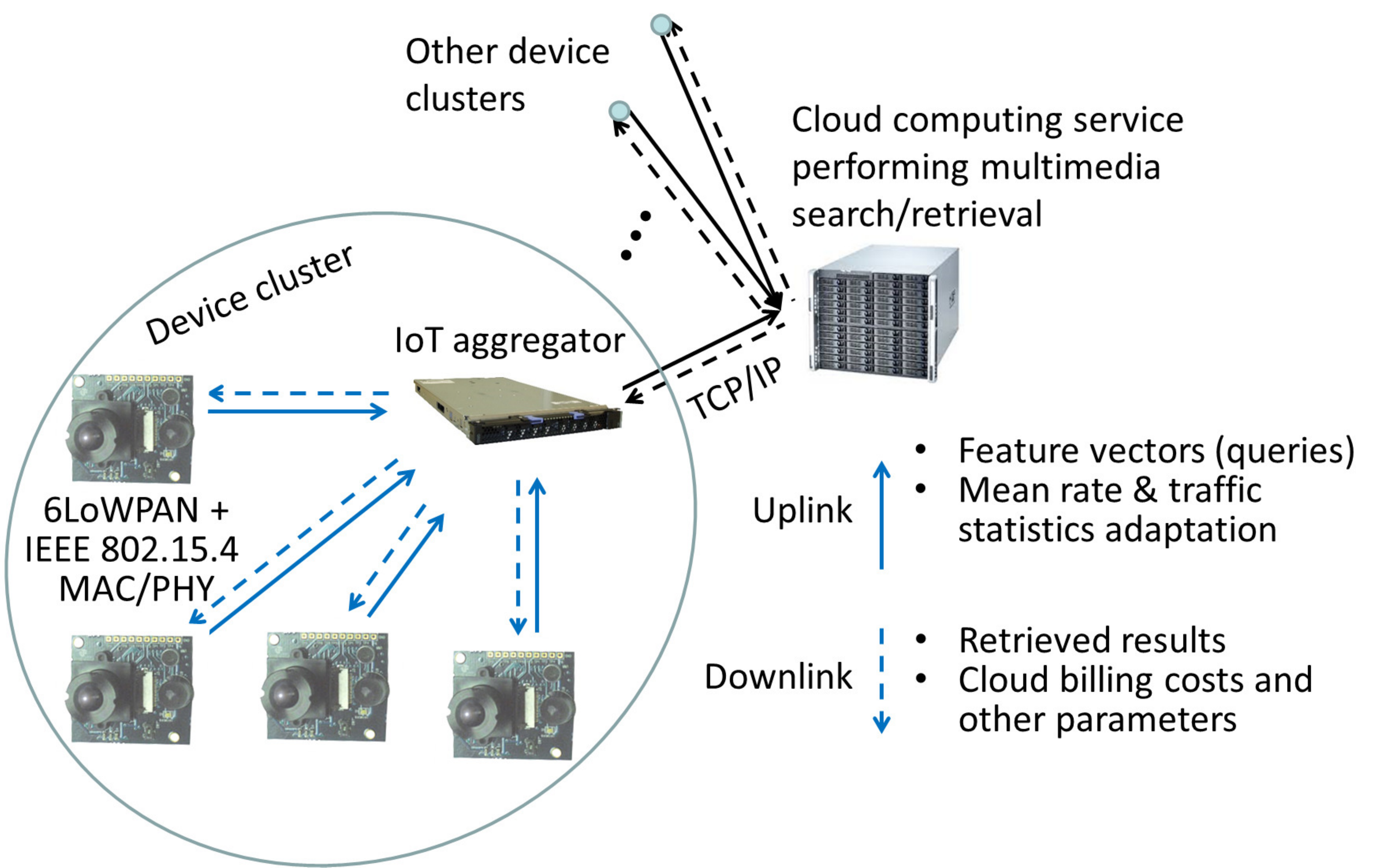} 
\par\end{centering}

\caption{System hierarchy for a media search application within an IoT context. Low-power devices send query data to an IoT aggregator using low-power protocols for the physical, medium access control and network layer, such as IEEE 802.15.4 MAC/PHY and 6LoWPAN. The IoT aggregator sends aggregated query volumes to the cloud-computing service using TCP/IP.  \label{fig:visual-search-system-diagram}}
\end{figure}

Figure \ref{fig:visual-search-system-diagram} presents an example
of how such applications can be deployed in practice within an Internet-of-Things (IoT) context. Energy-constrained devices capture and extract audio/visual features from audio and/or image streams and compact such features into feature-descriptor vectors \cite{arandjelovic2012three,perronnin2010large,jegou2012aggregating,serra2010audio}. Such feature vectors can be seen as \textit{queries} in a multimedia search application \cite{girod2011mobile,arandjelovic2012three}. For example, Serra \textit{et. al.} \cite{serra2010audio} propose beat and tempo feature extraction for cover song identification. A similar service is now widely deployed by Shazam. In the visual search domain,  several approaches produce image salient points and then compact their associated features into vectors of 64\(\sim\)8192 elements \cite{jegou2012aggregating,perronnin2010large}. All such feature vectors can be matched to equivalent vectors of very large content libraries via a cloud-based deployment within the context of classification, retrieval and similarity identification for, so-called, ``big data'' applications. {Devices of the same type running the same application software can be partitioned into ``device clusters'' (Fig. 1). Within each cluster, devices can be further subdivided into several classes based on the mean volume of query data produced within a certain time interval (e.g., ``low'', ``average'' and ``high'' volume of queries within a 24-hour interval). For instance, when cameras are deployed in a broad square, the data generation volume for some regions will be higher than others, depending on the expected activity of each region \cite{redondi2014energy}.} An IoT aggregator can be used to aggregate traffic from each device cluster and upload it to a remote cloud computing service that carries out the search operations for recognition and retrieval purposes \cite{5754008,6616112,6616113,ren2014dynamic}.

In this paper, we consider the energy consumption and billing costs incurred by such IoT applications in a holistic, system-oriented, manner. Specifically, we derive a parametric model that allows for the coupling of the energy consumption and cloud billing costs
in function of the system and query volume constraints for each cluster. 
A key aspect of our model is the derivation of the  \textit{optimal balancing} between:
\begin{enumerate}
\item
 \textit{idle time}, where device energy consumption or cloud billing cost is incurred for no useful output, e.g., image acquisition and processing or buffering/standby time on the device that does not lead to query generation, or cloud servers idling due to small volumes of queries being submitted;  
\item
\textit{active time}, where, despite resource consumption being incurred for useful output, one does not want to exceed certain limits in order not to cause excessive energy consumption in the device or excessive billing costs from the cloud infrastructure provider. 
\end{enumerate}
 Another key aspect of our work in comparison to previous
work on optimal energy management \cite{besbes2013analytic,redondi2014energy,beninidynamicpowermanagement,ZhangEnergyManagementspringer,alippienergymanagement,ren2014dynamic}
is that, beyond establishing the configuration that minimizes the device energy consumption metric of interest,\ we also derive closed-form expressions for the corresponding minimum
cloud billing cost, as well as the corresponding number of devices that can be admitted by the service.

In order to validate our analytic derivations, we utilize a proof-of-concept image similarity identification application, deployed via: \textit{(i)} running the image feature extraction and query generation and transmission on a Beaglebone Linux embedded platform; \textit{(ii)} implementing the back-end query processing for similarity identification and retrieval on Amazon Web Services Elastic Compute Cloud (AWS EC2) on-demand instances. Our results illustrate how the proposed model can be applied to real-world IoT-oriented media query retrieval systems in order to establish the desired operational parameters with respect to energy consumption and cloud infrastructure billing. {More broadly, the experimental results reported in this paper exemplify the efficacy of our framework for
feasibility studies on energy consumption and billing cost provisioning in cloud-based IoT query processing applications, prior to time-consuming testing and deployment.}   

The remainder of the paper is organized as follows. In \prettyref{sec:System Model}, we present the system model corresponding
to the application scenarios under consideration. The analytic derivations
characterizing energy-constrained feature extraction are presented in \prettyref{sec:Characterization-of-Energy},
where we also derive the optimal coupling with the utilized cloud-computing service
under four widely-used statistical characterizations for the query production rate. \prettyref{sec:Applications} presents
experimental results and \prettyref{sec:Conclusions} concludes
the paper.

\section{System Model\label{sec:System Model}}

Within the system hierarchy of Fig. \ref{fig:visual-search-system-diagram}, each   device connects to a ``repository''   service of a cloud provider, which represents the collecting unit, i.e. a cloud storage service like AWS Simple Storage Service (S3). This is where all device queries are uploaded to be processed by the back-end search mechanism of the service. As shown in Fig. \ref{fig:visual-search-system-diagram}, an IoT aggregator can be present in-between IoT clusters of the same type and the cloud repository, in order to: \textit{(i)} reshape the IoT query traffic volume before uploading it to the cloud; \textit{(ii)} carry out other device-specific and service-specific operations\footnote{Depending on the exact application, the IoT aggregator may carry out  authentication or encryption of queries, reformatting of the retrieved results from the cloud service so that they display correctly on the particular devices, application/collection of device metadata for service statistics and advertising, etc. We do not discuss these aspects as they are out of the scope of this paper. }. The figure shows that
the essentials of the problem boil down to the analysis of the interaction
between each mobile device node and its corresponding IoT aggregator and cloud computing service.

\subsection{System Description}

We assume that the mobile application is running continuously for a ``monitoring''\ interval of $T$
seconds. This interval corresponds to the typical device usage per day, or in-between battery recharging periods, e.g., $T \in \left[ 60,18000 \right]$ seconds per day. The activation, processing and transmission is either triggered by the user, or by external events at irregular times throughout the application's running time $T$. Examples are: user-triggered audio or visual feature extraction by recording a particular content segment (e.g., as in the Shazam, Google Voice or Google Goggles services), or motion-activated frame capturing and processing within an audio/visual surveillance application \cite{li2007streaming,anastasia2012throughput,spiliotopoulos2001quantization,andreopoulos2008incremental}. We therefore assume that the query data volume during $T$ seconds  is modeled as a random variable. 

The on-board battery of each device can last for prolonged intervals of time (e.g. tens or hundreds of hours, as it is typical in most audio/visual sensors and mobile devices). Therefore, the battery
capacity can be assumed to be infinite compared to the energy budget
spent by the media search application within each interval of $T$ seconds \cite{KansalSrivastava,yang2012optimal,tutuncuoglu2012optimum}. Hence, issues
such as leakage current and battery aging do not need to be considered. Nevertheless, it is well known that prolonged power surges from applications have inadvertent effects on a mobile device, such as faster-than-expected battery drainage or device overheating \cite{KansalSrivastava,tutuncuoglu2012optimum}. Therefore, in our analysis we shall be considering constraints on the mean energy consumption during the monitoring interval, as well as the one-sided deviation from the mean. Finally, we remark that the query data production and transmission and the cloud billing are not strictly continuous processes. However, given that we are focusing on large monitoring intervals comprising tens or hundreds of seconds, they can be seen as continuous processes.

\subsection{Definitions}

We now present the key concepts behind our analytic framework. The nomenclature summary of our system model is given in Table \ref{tab:Nomenclature-table.1}.

\subsubsection{Query data production}

The query data production and transmission by each device is
a non-deterministic process, because it depends on the frequency of the application invocation (or on event-driven activation alerts), as well as on the query size, which may vary, depending on the media search application. Therefore, the query data volume (in bits) for each time interval of $T$ seconds
of each device is modeled by random variable (RV) $\Psi_\mathrm{e}$ with probability density function (PDF) $P\left(\psi_\mathrm{e}\right)$. {More broadly, given that devices may be monitoring multiple event (or ``activity'') zones, e.g., low, medium and high activity, we consider $A$ activity zones with corresponding data volumes (in bits) for each time interval of $T$ seconds modelled by $A$ random variables with PDFs $P_{a}(\psi_{\mathrm{e}})$, $1\leq a\leq A$. 
The statistical modeling of data volumes for each activity zone} can be gained by observing the
occurred processing and analyzing the behavior of each device
when it captures image or audio data and produces query bits to be transmitted to the IoT\ aggregator. Alternatively, the query data production and transmission volume can be controlled (or ``shaped'') by the system designer in
order to achieve a certain goal, such as limiting the occurring latency or utilizing inactivity periods of other applications running concurrently on the mobile device. Examples of systems with variable
query data production and transmission rates include visual sensor networks transmitting
image features \cite{kulkarni2005senseye,rowe2007firefly,tagliasacchiWynerZivGaussian,redondilow},
as well as activity recognition networks where the
data acquisition is irregular and depends on the events occurring
in the monitored areas \cite{wernerallen2006fidelity,palma2010distributed,redondi2010laura}.

Beyond individual devices, the query volume uploaded from each IoT aggregator to  the cloud service is modelled by random variable  $\Psi_\mathrm{b}$ with PDF $P\left(\psi_\mathrm{b}\right)$. The distributions  $P_{a}\left(\psi_\mathrm{e}\right)$ and  $P\left(\psi_\mathrm{b}\right)$ will be of the same type (the latter will be a scaled version of the former) if the IoT\ aggregator shapes its uploaded traffic in the manner it receives it. 
Alternatively, if no traffic shaping is performed and the processing latency at the aggregator is fixed, {combining $n_1,\dots ,n_A$ devices producing queries from $A$ activity zones leads to:
\begin{equation}
P\left(\psi_\mathrm{b}\right)=\underset{n_1\:\mathrm{times}}{\underbrace{P_1\left(\psi_\mathrm{e}\right)\star\ldots\star P_1\left(\psi_\mathrm{e}\right)}}  \star\dots\star
\underset{n_A\:\mathrm{times}}{\underbrace{P_A\left(\psi_\mathrm{e}\right)\star\ldots\star P_A\left(\psi_\mathrm{e}\right)}}, 
\label{eq:P_b_convolution_P_e}
\end{equation}
\\
i.e., the PDF\ characterizing the uploaded traffic is the result of simple addition of the RVs modelling the data volumes received by all $n_\mathrm{tot}=\sum_{a=1}^An_a$ devices.}

Since the query data production volume may
be non-stationary, we assume its marginal statistics for $P_{a}\left(\psi_\mathrm{e}\right)$, $1\leq a \leq A$, and  $P\left(\psi_\mathrm{b}\right)$, which are derived starting from a doubly
stochastic model for these processes. Specifically, such marginal
statistics can be obtained by \cite{LamGoodmanDCT,foo2008analytical}:
\emph{(i)} fitting PDFs to sets of past measurements of query production volumes, with the statistical moments (parameters) of such distributions
characterized by another PDF; \emph{(ii)} integrating over the parameter
space to derive the final form of $P_{a}\left(\psi_\mathrm{e}\right)$ and  $P\left(\psi_\mathrm{b}\right)$.
For example, if the query data production is modeled as a Half-Gaussian
distribution with variance parameter that is itself exponentially
distributed, by integrating over the parameter space, the marginal
statistics of the query data volume become exponential \cite{LamGoodmanDCT,foo2008analytical}.
The disadvantage of using marginal statistics for the query data production
volume is the removal of the
stochastic dependencies to transient properties of these
quantities. However, in this work we are interested in constraining the first and second moments of the energy consumption, as well as minimizing the expected cloud billing cost, over
a lengthy time interval (e.g. several minutes or hours) and not in the instantaneous variations of these figures of merit  over short time intervals. Thus, a mean and variance-based
analysis using the marginal statistics is suitable for this purpose. 

\subsubsection{Energy and cloud infrastructure billing parameters}

We assume that, {on average}, the production and transmission of one query bit incurs energy consumption rate of $g_\mathrm{e}$ Joule-per-bit (J/b). This rate incorporates the audio or visual capturing, the feature extraction and compaction process to produce the compacted feature vector, and the transmission of the feature vector to the IoT\ aggregator. For example, under a visual search application, this would incorporate the energy consumption for the image acquisition, the processing to extract salient point descriptions, the compaction process to produce a 256-element feature vector comprising 32-bit numbers  (visual query) corresponding to each image \cite{jegou2012aggregating}, and the transceiver energy consumption to transmit this 8192-bit stream to the aggregator. Assuming that the entire process requires on average $10^{-5}$ J on the mobile device under consideration, this leads to $g_\mathrm{e} \cong 1.2 \times10^{-9}$ J/b. However, given the time-varying nature of the query data production per device,  we also encounter the case where the device is consuming energy to run the application (and possibly capture images or audio) in the background without producing any queries. This corresponds to ``idle'' energy consumption by each device with {average} rate $i_{\mathrm{e}}$ Joule-per-bit (i.e., $i_\mathrm{e}$ Joule for the time interval corresponding to the production and transmission of one query bit).  We assume that the application goes in idle mode during time intervals where the amount of produced query bits is below $c_\mathrm{e}E\left[\Psi_\mathrm{e}\right]$ b, with $E[\Psi_\mathrm{e}]$ the statistical expectation of  $\Psi_\mathrm{e}$. The value of $c_\mathrm{e}$ depends on the processing and transmission capabilities of the device, as well as on the specifics of the application, e.g., the size of the feature vector per query, the manner in which query generation is activated, etc. For instance, regular query generation (e.g., once per second) will correspond to lower value of $c_\mathrm{e}$ in comparison to motion-activated query generation, as the motion detection requires continuous capturing and processing of data that corresponds to higher percentage of ``idle'' energy consumption, i.e., energy consumption that does not lead to query data generation. {For small time intervals, the energy consumption rates $g_\mathrm{e}$ and $i_\mathrm{e}$ and the value for $c_\mathrm{e}$ may fluctuate depending on the device and network state (e.g., when memory paging, caching or other operating system tasks are carried out, or when transmission is hampered by high interference levels). However, given we are considering long periods of time for each monitoring interval $T$, we assume $g_\mathrm{e}$, $i_\mathrm{e}$ and $c_\mathrm{e}$ to represent the average values and our experimental validation demonstrates that accurate energy estimates can be derived under this assumption.}

Analogously, billing costs are incurred when servers are reserved from the cloud provider in order to process the queries uploaded by an IoT aggregator.  {Cloud providers offer a variety of usage-based pricing strategies for the consumption of computing resources that can be classified in three basic models: pay-as-you-go, subscription-based and auction-based. In this paper, we are primarily concerned with the first two models since auction-based models cannot guarantee reliable operation of compute instances.} 
 
{In the \textit{pay-as-you-go} model, users are billed with a static unit price per time interval without up-front costs or long-term commitments. Since the unit price remains constant, the total price increases linearly with the increase of the consumed units. Typically, a unit is a deployed virtual machine (e.g., an AWS\ EC2 instance) over a short time period (e.g., 1 hour) and the unit price follows a tiered model based on the compute capacity of the unit (in terms of CPU, memory and storage), the operating system and the region where the unit is deployed. For example the AWS EC2 ``on-demand'' instances use a linear per-unit pricing per hour. Variations of the linear pay-as-you-go model involve a step decrease in the unit price after certain utilization thresholds are reached, leading to a sublinear increase of the unit cost with the increase of usage time. An example of this model is the Sustained Usage pricing of the Google Cloud Engine.}  

{In \textit{subscription-based pricing} the user commits to long-term utilization for a pre-selected number of computing units by paying a fixed upfront price for the entire consumption period. Cloud providers may also offer a hybrid model that combines discounted pay-as-you-go pricing with an upfront payment for a fixed long-term time period (e.e 1 to 3 years). The AWS EC2 ``reserved'' instances and the Microsoft Azure Prepay implement a hybrid subscription/pay-as-you-go model. The unit price can be either linear or subject to a step decrease based on the usage volume.}

{Beyond their available pricing models,} all cloud computing services today use some form of autoscaling mechanism in order to adjust the number of compute instances according to the demand.  For example, in AWS Auto Scaling \cite{ryan2015aws} one can set rules that scale the utilized compute instances for every monitoring interval according to the average query volume received during the previous monitoring interval. A typical AWS Auto Scaling setup would be\footnote{The reported numbers of instances and instance types are only indicative and can be adjusted per IoT application.}: 
\begin{itemize}
\item
3 single-core AWS EC2 \texttt{m3.medium} on-demand instances when the average uploaded query volume was below a certain ``quota'' of $c_\mathrm{b}$ query bits (``idle'' case) in the previous monitoring interval,  

\item 30 on-demand instances when the query volume exceeded $c_\mathrm{b}$ b (``active'' case) in the previous monitoring interval. 
\end{itemize}
{Such configurations are prevalent in all cloud computing providers [AWS, Microsoft Azure, Google Compute Engine (GCE), etc.], where services are developed using a core number of instances, and additional compute instances are added when the demand exceeds a certain threshold \cite{ryan2015aws,li2010cloudcmp,6616112,6616113}.} For example, based on current AWS EC2 pricing, each single-core  \texttt{m3.medium}  instance incurs (on average) billing cost of 0.067\$ per hour under the  on-demand configuration. Assuming that a search operation with a $256\times 32$-bit query requires 10ms of cloud service time and under the AWS Auto Scaling rules stated above, this corresponds to billing cost of (approximately): $5.6\times10^{-7}$ dollars-per-query under the ``idle'' case, or $i_\mathrm{b} \cong 6.8 \times 10^{-11} $ dollars-per-query-bit (\$/b) and  $p_\mathrm{b}\cong 6.8 \times 10^{-10} $ \$/b for the ``active'' case. {Similar billing rates can be calculated for other cloud providers under pay-as-you-go or subscription-based pricing. Notably, despite the fact that cloud infrastructure billing is levied on hourly or minute-by-minute increments (e.g., for AWS and GCE, resp.), because of the continued nature of the service, we do not have ``termination gaps''. Instead, some instances may idle for some time before they are reused or terminated, depending on the fluctuations of the query volume within each monitoring interval. The quota of $c_\mathrm{b}$ query bits that triggers the auto scaling can be set according to the application or the number of devices within each IoT aggregator and the billing rates are always linear to the number of queries since we always consider a fixed query and database size, which leads to the computation time increasing linearly to the number of queries.} 

{Beyond the cost of the computing time, billing cost proportional to the expected query volume per monitoring interval, $E[\Psi_\mathrm{b}]$, must be constrained to $V_\mathrm{max}$ b, since: \textit{(i)} all cloud providers charge for data transfers and storage; and \textit{(ii)}\ excessive interference will occur if the average query volume rises above the capacity of the local network of each IoT\ aggregator.} Assuming 0.15\$\  per gigabyte of query volume (based on current AWS pricing), this leads to (approximately) $g_\mathrm{b}=1.9\times 10^{-11}$ \$/b. \ Then, in order to remain competitive against other solutions in the market, the service may wish to set an expectation that each user should be billed for $B_\mathrm{mean}$ \$ on average for each device and each monitoring time interval of $T$ seconds. {Importantly, while the instantaneous rates $i_\mathrm{b}$, $p_\mathrm{b}$, $g_\mathrm{b}$ and the instantaneous query transmission rate from each IoT\ aggregator may fluctuate, given that we are interested in long monitoring intervals and infrastructure billing is typically applied in minute or even hourly increments, we utilize mean values for these rates, calculated by averaging over lengthy operational periods. }

Evidently, the large number of system, data production,  energy consumption, and cloud billing parameters of Table \ref{tab:Nomenclature-table.1} makes the exhaustive exploration of the complete design space infeasible. Therefore, the creation of an analytic model that can establish closed-form relationships between the different parameters, as well as optimal settings under specified conditions for device energy consumption and billing cost is of paramount importance. This is the aim of the next section.

\begin{table}[tb]
\noindent \centering{}\caption{\label{tab:Nomenclature-table.1}Nomenclature table.}
\begin{tabular}{>{\centering}m{0.2\columnwidth}>{\centering}m{0.02\columnwidth}>{\raggedright}p{0.65\columnwidth}}
\multicolumn{1}{>{\centering}m{0.2\columnwidth}}{Symbol } & \multicolumn{1}{>{\centering}m{0.02\columnwidth}}{Unit} & \multicolumn{1}{>{\raggedright}m{0.65\columnwidth}}{Definition}\tabularnewline
\toprule
$n_\mathrm{tot},n_{a}$ & -- & {Total number of devices and devices of activity zone $a$ (out of $A$ total zones) within the same IoT aggregator} \tabularnewline
\midrule
$g_{\mathrm{e}}$ & J/b & Energy for producing and transmitting a query bit\tabularnewline
\midrule
$i_{\mathrm{e}}$ & J/b & Energy during idle periods equal to the interval required to produce and transmit a query bit\tabularnewline
\midrule
$c_{\mathrm{e}}$ & --  & Fraction of the average query volume below which the device application is in idle mode\tabularnewline
\midrule
$E_\mathrm{max\_exp}$ & J & Upper bound of the expected energy consumption over $T$ seconds\ \tabularnewline
\midrule
$E_\mathrm{max\_var}$ & J$^2$ & Upper bound of the one-sided variation from the expected energy consumption over $T$ seconds\ \tabularnewline
\midrule
$r_\mathrm{tot},r_{a}$ & {b} & {Average query data production and transmission volume and average volume per device of activity zone $a$ (over the monitoring interval)}\tabularnewline
\midrule
{$V_\mathrm{max}$} & {b} & {Maximum query data transmission volume of each IoT aggregator over the monitoring interval}\tabularnewline
\midrule
$\Psi_\mathrm{e}\sim P_{a}\left(\psi_\mathrm{e}\right)$, $\Psi_\mathrm{b}\sim P\left(\psi_\mathrm{b}\right)$ & b & {RVs modeling the query data production and transmission volume per device (and activity zone $a$)\ and per IoT aggregator}\tabularnewline
\midrule
$E\left[\Psi_\mathrm{e}\right]$, $E\left[\Psi_\mathrm{b}\right]$ & b & Expected query data production and transmission  per device and per aggregator over the monitoring time interval\tabularnewline
\midrule
$g_{\mathrm{b}}$ & \$/b & Billing cost (per query bit) incurred from uploading/storing a query\tabularnewline
\midrule
$i_{\mathrm{b}}$ & \$/b & Billing cost (per time interval corresponding to the processing time per query bit) incurred from ``idle'' periods \tabularnewline
\midrule
$c_{\mathrm{b}}$ & b & Number of query bits (quota)\ above which the cloud Auto Scaling mechanism switches from idle to active state
\tabularnewline
\midrule
$p_{\mathrm{b}}$ & \$/b & Billing cost (per query bit) incurred from processing a query after exceeding the quota of $c_\mathrm{b}$ query bits per $T$ seconds (``active'' period)\tabularnewline
\midrule
$B_{\mathrm{mean}}$ & \$ & Expected cloud billing cost over $T$ seconds\tabularnewline\bottomrule
\end{tabular}
\end{table}

\section{Characterization of Energy Consumption and Cloud Billing Cost\label{sec:Characterization-of-Energy}}

We derive analytic expressions for the expected energy consumption of a device (and its one-sided deviate), as well as the expected cloud billing\ for {a group of $n_\mathrm{tot}$ devices on the same IoT aggregator}. This allows us to derive closed-form conditions that ensure that the one-sided energy variation is minimized under a constraint on the  expected energy consumption for each device, or, vice-versa. We also derive the conditions that minimize the incurred billing cost and ensure that the minimum value can be set to the expected billing of $B_\mathrm{mean}$ per monitoring period of $T$ seconds, {while satisfying the total query transmission volume constraint, $V_{\mathrm{max}}$, of the IoT aggregator.}  

The expected energy consumption of each mobile device {of activity zone $a$} over the monitoring period of $T$ seconds is: 
\begin{equation}
E_{\mathrm{exp}}  =  E\left[\Psi_\mathrm{e}\right]g_{\mathrm{e}}  +  i_\mathrm{e}\int_{0}^{c_\mathrm{e} E\left[\Psi_\mathrm{e}\right] }\left(c_\mathrm{e} E\left[\Psi_\mathrm{e}\right]-\psi_\mathrm{e}\right)P_{a}\left(\psi_\mathrm{e}\right)d\psi_\mathrm{e}, \label{eq:E_exp}
\end{equation}
where the integral of the second term expresses the expected energy consumption for the time that the device will be in idle mode. {This term expresses the energy consumed to produce no useful output, i.e., energy consumed that does not lead directly to the production of query volume (e.g., image acquisition and processing or buffering/standby).}

 We can also express the one-sided variability of the energy consumption when the application switches from idle to active state (i.e., when exceeding the $c_\mathrm{e} E\left[\Psi_\mathrm{e}\right]$-bit query volume):
\begin{equation}
E_{\mathrm{var}}  =  g_{\mathrm{e}}^2\int_{c_{\mathrm{e}}E\left[\Psi_\mathrm{e}\right]}^{\infty}\left(\psi_\mathrm{e} - c_\mathrm{e}E\left[\Psi_\mathrm{e}\right]\right)^2 P_{a}\left(\psi_\mathrm{e}\right)d\psi_\mathrm{e}. 
\label{eq:E_var}
\end{equation}
 {For each monitoring interval of $T$ seconds, higher values of $E_{\mathrm{var}}$ imply higher energy consumption fluctuation from the average energy consumption.}
Therefore, under a given energy budget of $E_\mathrm{exp}$ Joule for the monitoring time interval of $T$ seconds, allowing for a large value for $E_\mathrm{var}$ will incur significant drop in the device battery level (and possibly other unintended consequences, such as device overheating, battery degradation over time, etc.).  On the other hand, a small value of $E_\mathrm{var}$ will limit the query production volume handled by the device, or may require a very high value for $c_\mathrm{e}$ that may not be realistic for the application and hardware under consideration.


Let us now consider the expected cloud billing cost {when receiving $n_\mathrm{tot}$  aggregated media query volumes from an IoT\ aggregator}. We can express this cost via 
\begin{eqnarray}
B_{\mathrm{exp}} & = &    E\left[\Psi_\mathrm{b}\right]g_\mathrm{b} + i_\mathrm{b}\int_{0}^{c_\mathrm{b}}\left(c_\mathrm{b}-\psi_\mathrm{b}\right)P\left(\psi_\mathrm{b}\right)d\psi_\mathrm{b}\nonumber \\ 
& + &  p_\mathrm{b}\int_{c_\mathrm{b}}^{\infty}\left(\psi_\mathrm{b} - c_\mathrm{b}\right)P\left(\psi_\mathrm{b}\right)d\psi_\mathrm{b}, \label{eq:B_exp definition}
\end{eqnarray}
where: $E\left[\Psi_\mathrm{b}\right]g_\mathrm{b}$ corresponds to the data transfer/storage costs, the first integral corresponds to the partial moment expressing the ``idle'' billing cost, and the second integral corresponds to the ``active'' billing. Adding and subtracting $p_{\mathrm{b}}\int_{0}^{c_\mathrm{b}}\left(\psi_\mathrm{b}-c_{\mathrm{b}}\right)P\left(\psi_\mathrm{b}\right)d\psi_\mathrm{b}$
in $B_{\mathrm{exp}}$, we get: 
\begin{eqnarray}
B_{\mathrm{exp}} & = & E\left[\Psi_\mathrm{b}\right]\left(g_\mathrm{b} + p_\mathrm{b} \right)-p_\mathrm{b}c_\mathrm{b} \nonumber \\ 
& + &  \left(i_\mathrm{b}+p_\mathrm{b} \right)\int_{0}^{c_\mathrm{b}}\left(c_\mathrm{b}-\psi_\mathrm{b}\right)P\left(\psi_\mathrm{b}\right)d\psi_\mathrm{b}. 
\label{eq:B_exp simplified}
\end{eqnarray}

Evidently, the expected billing cost depends on the coupling point, $c_\mathrm{b}$,
as well as on the PDF of the aggregate query data reaching the cloud service,
$P\left(\psi_\mathrm{b}\right)$, which is either a variant of the  $P_{a}\left(\psi_\mathrm{e}\right)$ distributions, or it is linked to them via \eqref{eq:P_b_convolution_P_e}. In the remainder of this section: 
\begin{itemize}
\item
We consider various cases for $P_{a}\left(\psi_\mathrm{e}\right)$ and $P\left(\psi_\mathrm{b}\right)$ and minimize the energy variance of \eqref{eq:E_var}  subject to an upper bound for the expression of \eqref{eq:E_exp}, and  vice-versa.
\item
We derive the number of query bits (quota), $c_\mathrm{b}$,\ that  minimizes the corresponding billing cost of \eqref{eq:B_exp simplified} under various PDFs, $P\left(\psi_\mathrm{b}\right)$. 
\item
In order for the desired energy consumption and billing cost parameters to be met concurrently {while obeying the total traffic volume constraint of each IoT aggregator}, we associate the minimum billing cost with the desired value for the expected billing, $B_\mathrm{mean}$, and the device query production volumes {for  activity zone. Therefore, we establish the corresponding number of devices, $n_1,\dots n_A$, that can be admitted by each IoT\ aggregator under the optimal configuration.}
\end{itemize}

\subsection{Coupling of Device Energy Consumption and Cloud Infrastructure Billing}

In order to control the overall energy consumption profile of the application, one may wish to minimize the expected one-sided energy variability,  $E_\mathrm{var}$, subject to the constraint that the expected energy consumption  does not exceed   $E_{\mathrm{max\_exp}}$ Joule within $T$ seconds. Both of these values are provided by the application or device developer in order to ensure the application does not degrade the user quality-of-experience, or disrupt other concurrently-running services on the device.
We term this problem as the ``primary'' optimization problem, and its converse as the ``dual'' problem. 

\subsubsection{Primary energy minimization problem}

We  determine the value $c_{\mathrm{e}}$ that minimizes the one-sided variability of the energy consumption while satisfying a constraint on the average energy consumption, i.e., we consider the optimization problem
\begin{equation}
\begin{aligned}
& \underset{c_{\mathrm{e}} \in \mathbb{R}^+}{\text{minimize}}
& &  E_{\mathrm{var}}\\
& \text{subject to}
& & E_{\mathrm{exp}} \leq E_{\mathrm{max\_exp}}.
\end{aligned}
\label{eq:primal}
\end{equation}

\subsubsection{Dual energy minimization problem}

Consider now a dual setting, in which one aims at minimizing the average energy consumption while satisfying a constraint on the maximum one-sided energy variation from idle to active mode. The activation threshold $c_{\mathrm{e}}$ that achieves this is found by solving the optimization problem
\begin{equation}
\begin{aligned}
& \underset{c_{\mathrm{e}} \in \mathbb{R}^+}{\text{minimize}}
& &   E_{\mathrm{exp}} \\
& \text{subject to}
& &E_{\mathrm{var}} \leq E_{\mathrm{max\_var}}.
\end{aligned}
\label{eq:dual}
\end{equation}

\subsubsection{Convexity of the energy minimization problems and closed-form solutions}

We first show that both the primary and dual optimization problems of \eqref{eq:primal} and \eqref{eq:dual} are convex. Therefore, they can be solved using fast numerical methods, such as gradient descent or the Newton-Raphson method. 

By taking the first and the second derivative of $E_{\mathrm{exp}}$ with respect to $c_{\mathrm{e}}$ we obtain
\begin{IEEEeqnarray}{rCl}
\frac{d E_{\mathrm{exp}}}{d c_{\mathrm{e}}}  &  = &  i_{\mathrm{e}} E[\Psi_{\mathrm{e}}] F _{a}(c_{\mathrm{e}} E[\Psi_{\mathrm{e}}]  )  \\
\frac{d^2 E_{\mathrm{exp}}}{d c_{\mathrm{e}}^2}  & = & i_{\mathrm{e}} (E[\Psi_{\mathrm{e}}])^2 P_{a} (c_{\mathrm{e}} E[\Psi_{\mathrm{e}}]  ),
\end{IEEEeqnarray}
where $F_{a}(\psi_{\mathrm{e}})$ and $P_{a}(\psi_{\mathrm{e}})$ are the cumulative distribution function (CDF) and the PDF of the query volume per device of activity zone $a$, $\Psi_{\mathrm{e}}$, respectively. Since $\frac{d^2 E_{\mathrm{exp}}}{d c_{\mathrm{e}}^2} \geq 0$,  $E_{\mathrm{exp}}$ is a convex function of $c_{\mathrm{e}}$.

Analogously, by taking the first and the second derivative of $E_{\mathrm{var}}$ with respect to $c_{\mathrm{e}}$ we obtain
\begin{IEEEeqnarray}{rCl}
\nonumber
\frac{d E_{\mathrm{var}}}{d c_{\mathrm{e}}}  &  = &  2 g_{\mathrm{e}}^2 (E[\Psi_{\mathrm{e}}])^2 c_{\mathrm{e}} \left[   1- F_{a} (c_{\mathrm{e}} E[\Psi_{\mathrm{e}}]  )  \right]\\
&&  - 2 g_{\mathrm{e}}^2  E[\Psi_{\mathrm{e}}]  \int_{c_{\mathrm{e}}  E[\Psi_{\mathrm{e}}]}^{+\infty} \psi_{\mathrm{e}}  P_{a}(\psi_{\mathrm{e}}) d \psi_\mathrm{e}  \\
\frac{d^2 E_{\mathrm{var}}}{d c_{\mathrm{e}}^2}  & = & 2 g_{\mathrm{e}}^2(E[\Psi_{\mathrm{e}}])^2 \left[   1- F_{a} (c_{\mathrm{e}} E[\Psi_{\mathrm{e}}]  )  \right].
\end{IEEEeqnarray}
Thus,  $\frac{d^2 E_{\mathrm{var}}}{d c_{\mathrm{e}}^2} \geq 0$ and $E_{\mathrm{var}}$ is also a convex function of $c_{\mathrm{e}}$.

Given that $E_\mathrm{exp}$ and $E_\mathrm{var}$ are convex, the following proposition offers a way to derive the solutions of the problems \eqref{eq:primal} and \eqref{eq:dual} in closed form.
\begin{prop}
\label{prop:monotonicity}
The solution to the optimization problem \eqref{eq:primal} is such that, at the optimal $c_{\mathrm{e}}$, it holds
\begin{equation}
E_{\mathrm{exp}} = E_{\mathrm{max\_exp}}.
\label{eq:primal_sol}
\end{equation}
The solution to the optimization problem \eqref{eq:dual} is such that, at the optimal $c_{\mathrm{e}}$, it holds
\begin{equation}
E_{\mathrm{var}} = E_{\mathrm{max\_var}}.
\label{eq:dual_sol}
\end{equation}
\end{prop}
\begin{IEEEproof}
See Appendix.
\end{IEEEproof}
In other words, the solutions to both optimization problems are obtained when the constraints are met with equality. Therefore, whenever possible, by inverting the closed-form expressions of $E_{\mathrm{exp}}$ and $E_{\mathrm{var}}$ for different query volume PDFs, we can find the optimal $c_{\mathrm{e}}$ in closed form.

\subsubsection{Billing parameter tuning to minimize the cloud infrastructure billing cost and meet the expected billing $B_\mathrm{mean}$}

We can now turn our attention to the billing cost $B_\mathrm{exp}$ in \eqref{eq:B_exp simplified} for the $n_\mathrm{tot}$-device aggregate query production volume over the monitoring time interval of $T$ s. We note that the first and the second derivative of $B_{\mathrm{exp}}$ with respect to the coupling point $c_{\mathrm{b}}$ are given by
\begin{IEEEeqnarray}{rCl}
\frac{d  B_{\mathrm{exp}}}{d c_{\mathrm{b}}}  & = &  - p_{\mathrm{b}}  + (i_{\mathrm{b}}   + p_{\mathrm{b}}) F ( c_{\mathrm{b}})  \\
 \frac{d^2  B_{\mathrm{exp}}}{d c_{\mathrm{b}}^2}  & = & (i_{\mathrm{b}}   + p_{\mathrm{b}}) P ( c_{\mathrm{b}}) ,
\end{IEEEeqnarray}
where $F ( \psi_{\mathrm{b}})$ and $P ( \psi_{\mathrm{b}})$ are CDF and the PDF of the aggregated query volume $\Psi_{\mathrm{b}}$. Therefore, we can conclude that $B_{\mathrm{exp}}$ is a convex function of $c_{\mathrm{b}}$ when $\Psi_{\mathrm{b}}$ is modelled by a continuous distribution function. Moreover, the value of $c_{\mathrm{b}}$ that minimizes the billing cost is obtained by solving the equation $\frac{d  B_{\mathrm{exp}}}{d c_{\mathrm{b}}}=0$, i.e.,
\begin{equation}
c_{\mathrm{b}}  = F^{-1} \left(  \frac{p_{\mathrm{b}}}{ i_{\mathrm{b}} + p_{\mathrm{b}} }  \right),
\label{eq:cb_opt_general}
\end{equation}
where $F^{-1} (\cdot)$ is the inverse CDF of $\Psi_{\mathrm{b}}$. Assuming any strictly-increasing CDF, $c_\mathrm{b}$ will be unique\footnote{Even if the CDF is monotonically increasing, all candidate extrema are equivalent with respect to the derived billing cost.}. Therefore,
in conjunction with the fact that $\forall c_\mathrm{b}: \frac{d^{2}  B_{\mathrm{exp}}}{d c^2_{\mathrm{b}}}>0$, $B_\mathrm{exp}$ attains a global minimum in function of $c_\mathrm{b}$.

\subsubsection{Number of devices in an IoT aggregator to concurrently satisfy cost and system constraints}

In order to meet energy, billing and query volume constraints: $\left\{E_{\mathrm{max\_exp}}\text{ or }E_\mathrm{max\_var}\right\} $, $B_\mathrm{mean}$ and $V_\mathrm{max}$, we first find $c_{\mathrm{e}}$ corresponding to \eqref{eq:primal_sol} or \eqref{eq:dual_sol}. We can then match {the device query volumes $r_1,\dots, r_A$} with the minimum billing cost, $\min \{ B_{\mathrm{exp}} \}$, obtained by substituting $c_\mathrm{b}$ from \eqref{eq:cb_opt_general} into \eqref{eq:B_exp simplified}. Finally, setting 
\begin{equation}
\min \{ B_{\mathrm{exp}} \} = B_{\mathrm{mean}},
\end{equation}
{and constraining the average traffic volume of all devices,
\begin{equation}
r_\mathrm{tot}=\sum_{a=1}^A{n_ar_a}, \label{eq:r_tot}
\end{equation}
by
\begin{equation}
r_\mathrm{tot}  \leq V_{\mathrm{max}},
\end{equation}
}
\\
we obtain the number of devices, $n_1,\dots ,n_A$, that can be accommodated by an IoT aggregator when each device satisfies the energy settings of \eqref{eq:primal} or \eqref{eq:dual} and the IoT-uploaded volume incurs the minimum billing cost of $B_\mathrm{mean}$ \$ per monitoring interval, {while satisfying the query volume constraint $V_{\mathrm{max}}$}.

Overall, via the energy-constrained analysis and the cloud-billing optimization, one can explore different energy and billing settings in order to accommodate particular types of mobile devices (with given energy consumption parameters), predetermined average query production volume, or given number of devices per IoT cluster of Fig. \ref{fig:visual-search-system-diagram}. We present detailed examples for this in the following subsections.   

\subsection{Illustrative Case: $\Psi_\mathrm{e}$ and $\Psi_\mathrm{b}$ Are Uniformly Distributed\label{sub:Uniform Case}}

{For each activity zone $a$,} when no knowledge of the underlying statistics of the query generation
process exists, one can assume that both $P_{a}\left(\psi_\mathrm{e}\right)$ and $P\left(\psi_\mathrm{b}\right)$ are
uniform over the intervals {$\left[0,2r_{a}\right]$ and $\left[0,2 r_{\mathrm{tot}}\right]$, respectively: 
\begin{equation}
P_{\mathrm{U},a}\left(\psi_\mathrm{e}\right)=\left\{ \begin{array}{c}
\frac{1}{2r_{a}},\\
0,
\end{array}\begin{array}{c}
0\leq\psi_\mathrm{e}\leq2r_{a}\\
\mathrm{otherwise}
\end{array},\right.\label{eq:P_U_psi_e_uniform}
\end{equation}
and

\begin{equation}
P_{\mathrm{U}}\left(\psi_\mathrm{b}\right)=\left\{ \begin{array}{c}
\frac{1}{2r_{\mathrm{tot}}},\\
0,
\end{array}\begin{array}{c}
0\leq\psi_\mathrm{b}\leq2r_{\mathrm{tot}}\\
\mathrm{otherwise}
\end{array}.\right.\label{eq:P_U_psi_b_uniform}
\end{equation}}
This corresponds to the case where the IoT aggregator's upload query volume PDF matches the query generation PDF \eqref{eq:P_U_psi_e_uniform} and {the aggregator merges and transmits query volumes of {$n_\mathrm{tot}$} devices to the cloud service under the query volume PDF of \eqref{eq:P_U_psi_b_uniform}.} 

{For each activity zone $a$, $1\leq a \leq A$, the expected value of $\Psi_\mathrm{e}$ is $E_{\mathrm{U}}\left[\Psi_\mathrm{e}\right]=r_a$ b.} The expected value of  $\Psi_\mathrm{b}$ is $E_{\mathrm{U}}\left[\Psi_\mathrm{b}\right]=r_{\mathrm{tot}}$
b. The cases where $c_{\mathrm{e}}>2$ or $c_{\mathrm{b}}>2r_{\mathrm{tot}}$ are of no practical relevance, because:\ \textit{(i)} the first inequality means each device is always in idle mode, or \textit{(ii)} the second inequality means the cloud infrastructure is constantly overprovisioned. Thus, we are only concerned with the case where:  $0<c_{\mathrm{e}}<2$ and $0<c_{\mathrm{b}}<2r_{\mathrm{tot}}$. 
\subsubsection{Energy parameter tuning corresponding to the solution of the Primary and Dual minimization problems of \eqref{eq:primal} and \eqref{eq:dual}}
Starting from the device energy consumption, by using \eqref{eq:P_U_psi_e_uniform} in \eqref{eq:E_exp}, 
we obtain:
\begin{equation}
E_{\mathrm{exp,U}}  =  \left(g_{\mathrm{e}}  + \frac{ i_\mathrm{e} c^2_{\mathrm{e}}}{4} \right)r_{a}. 
\label{eq:E_exp_uniform}
\end{equation}
In addition, by using \eqref{eq:P_U_psi_e_uniform} in \eqref{eq:E_var},
we obtain: 
\begin{equation}
E_{\mathrm{var,U}}  =  g_{\mathrm{e}}^2 \frac{\left(2-c_\mathrm{e}\right)^3}{6}r_{a}^2.
 \label{eq:E_var_uniform}
\end{equation}

%
%

Then, given the average query volume $r_{a}$ per time interval $T$, and the corresponding energy parameters $i_\mathrm{e}$ and $g_{\mathrm{e}}$, it is possible to derive the activation threshold $c_{\mathrm{e}}$ that corresponds to the solution to \eqref{eq:primal} by solving $E_{\mathrm{exp}} = E_{\mathrm{max\_exp}}$ for $c_{\mathrm{e}}$. Thus, we obtain
\begin{equation}
c_\mathrm{e,U,primary} = 2 \sqrt{\frac{E_{\mathrm{max\_exp}}  - g_{\mathrm{e}}r_{a}  }{i_{\mathrm{e}}r_{a}}},
\label{eq:c_e_Uniform}
\end{equation}
provided that $E_{\mathrm{max\_exp}}>g_{\mathrm{e}} r_{a}$. The last inequality must hold or else the energy constraint does not suffice for the production of $r_{a}$ b within $T$ seconds.  We also note that the constraint $c_{\mathrm{e}}<2$ implies in this case that $E_{\mathrm{max\_exp}}< (g_{\mathrm{e}}+ i_{\mathrm{e}}) r_a$. These two constraints provide the feasible range for the expected energy consumption under Uniformly-distributed query volumes as: $E_{\mathrm{max\_exp}}\in \left( g_{\mathrm{e}} r, (g_{\mathrm{e}}+ i_{\mathrm{e}}) r _{a}\right)$.  


Similarly, the solution to the constrained miminization of \eqref{eq:dual} is obtained by solving for $c_{\mathrm{e}}$ the equation $E_{\mathrm{var}}  = E_\mathrm{max\_var}$, thus obtaining
\begin{equation}
c_{\mathrm{e,U,dual}} = 2 - \left(  \frac{6 E_\mathrm{max\_var}}{g_{\mathrm{e}}^2 r_{a}^2}    \right)^{1/3}.
\label{eq:c_e_Uniform_2}
\end{equation}
Note that if the constraint on the one-sided energy variation is such that $E_\mathrm{max\_var} \geq 4 g_{\mathrm{e}}^2 r_{a}^2/3 $, then such constraint is verified by all nonnegative values of $c_{\mathrm{e}}$, and the minimum average energy consumption is achieved by setting $c_{\mathrm{e}}=0$.
Due to the finite support of the Uniform distribution, this effectively corresponds to the trivial case when the one-sided deviation is unlimited and the minimum energy consumption is obtained when no idle energy is consumed.    
\subsubsection{Billing parameter tuning to minimize the cloud infrastructure billing cost and meet the expected billing $B_\mathrm{mean}$}

For the case of uniform distribution, by replacing  \eqref{eq:P_U_psi_b_uniform} in \eqref{eq:B_exp simplified}, we obtain the average billing cost as
{
\begin{equation}
B_{\mathrm{exp,U}}  =  \left(g_\mathrm{b} + p_\mathrm{b} \right)r_{\mathrm{tot}}-p_\mathrm{b}c_\mathrm{b}  +  \left(i_\mathrm{b}+p_\mathrm{b} \right)\frac{c^2_\mathrm{b}}{4r_{\mathrm{tot}}}, 
\label{eq:B_exp_uniform}
\end{equation}}
and the optimal coupling point \eqref{eq:cb_opt_general} is
{
\begin{equation}
c_{\mathrm{b,U}}=\frac{ 2p_\mathrm{b}r_{\mathrm{tot}}}{i_\mathrm{b}+p_\mathrm{b}}.
\label{eq:c_b,U}
\end{equation}}
The corresponding minimum-possible billing cost is:
{
\begin{equation}
\mathrm{min}\left\{ B_\mathrm{exp,U}\right\}   =  \left(g_\mathrm{b} + p_\mathrm{b} -\frac{p^2_\mathrm{b}}{i_\mathrm{b}+p_\mathrm{b}}\right)r_{\mathrm{tot}}.
\label{eq:min_B_exp_U}
\end{equation}}
The last equation shows that the minimum billing cost increases linearly to the average query data production volume of all {$n_\mathrm{tot}$} devices. 

{\subsubsection{Number of devices in an IoT aggregator to concurrently satisfy cost and system constraints}
In order to meet \textit{both} energy and billing costs: $\left\{E_{\mathrm{max\_exp}},E_\mathrm{max\_var}\right\} $ and $B_\mathrm{mean}$, we can first tune $c_{\mathrm{e}}$ according to \eqref{eq:c_e_Uniform} or \eqref{eq:c_e_Uniform_2}. Then, by substituting $\min \{ B_{\mathrm{exp,U}} \}= B_{\mathrm{mean}}$ in \eqref{eq:min_B_exp_U}, the expected billing $B_{\mathrm{mean}}$ is achievable under the query transmission volume constraint $V_{\mathrm{max}}$  if 
\begin{equation}
B_{\mathrm{mean}} \leq  {V_{\mathrm{max}} }{ \left(g_\mathrm{b} + p_\mathrm{b} -\frac{p^2_\mathrm{b}}{i_\mathrm{b}+p_\mathrm{b}}\right)}.
\label{eq:feasibility_Bmean}
\end{equation}
Otherwise, $V_\mathrm{max}$ cannot accommodate the query volume that guarantees billing equal to $B_{\mathrm{mean}}$. When \eqref{eq:feasibility_Bmean} is satisfied, we can use \emph{proportional fairness} \cite{kelly1998rate} to derive the number of devices from different activity zones, i.e.,
\begin{equation}
n_{\mathrm{U},a} =\frac{B_\mathrm{mean}}{\left(g_\mathrm{b} + p_\mathrm{b} -\frac{p^2_\mathrm{b}}{i_\mathrm{b}+p_\mathrm{b}}\right)A r_{a}}, 
\label{eq:n_Ua}
\end{equation}
for $1\leq a \leq A$. An interesting solution for $n_{\mathrm{U},a}$ occurs if $B_\mathrm{mean}$ is set so that the volume  $V_\mathrm{max}$ is expected to be fully utilized, i.e., the constraint of \eqref{eq:feasibility_Bmean} is met with equality. In such a case, the energy consumption parameters ($E_\mathrm{exp,U}$ and $E_\mathrm{var,U}$), the desired cloud billing cost ($B_\mathrm{mean}$), and the aggregator's data transmission volume ($V_\mathrm{max}$) become \textit{mutually coupled}. Then, the number of devices from different activity zones that can accommodated by the IoT aggregator simply becomes
\begin{equation}
n_{\mathrm{U},a}= \frac{V_\mathrm{max}}{Ar_{a}}.
\label{eq:prop_fair}
\end{equation}
}
\\Overall,  under the uniform distributions of \eqref{eq:P_U_psi_e_uniform} and \eqref{eq:P_U_psi_b_uniform}, {$n_{\mathrm{U},a}$ of \eqref{eq:n_Ua}  represents the number of devices that should be accommodated by an IoT aggregator [with each device having $c_{\mathrm{e}}$ according to \eqref{eq:c_e_Uniform} or \eqref{eq:c_e_Uniform_2}] in order to lead to the minimum billing cost being equal to $B_\mathrm{mean}$ and aggregated query volume below or equal to $V_\mathrm{max}$ b.}

\subsection{Energy-constrained Query Volume Production and Minimum Billing Cost under Pareto, Exponential and Half-Gaussian Distributions\label{sub:Minimum-Expected-Power-OtherPDF}}

We can now extend the previous calculation to other distributions
expressing commonly observed data transmission rates in practical
applications. We consider three additional PDFs for $\Psi_{\mathrm{e}}$ and $\Psi_{\mathrm{b}}$ that have
been used to model the marginal statistics of many real-world data
transmission applications and provide the obtained analytic results
in this subsection. For each distribution {and for each activity zone $a$}, we couple its parameters to the average
query volume of the uniform distribution, $r_a$.
This facilitates comparisons of the energy consumption and billing cost achievable under different
statistical characterizations for the query volume.

\subsubsection{Pareto distribution and fixed query volume}

This distribution has been used, amongst others, to model the marginal
data size distribution of data production processes that result in substantial number
of small data volumes and a few very large ones \cite{paxsonTCP,parkTCP}. 
{For each activity zone $a$, $1 \leq a \leq A$}, consider $P_{\mathrm{P},a}\left(\psi_{\mathrm{e}} \right)$ as the Pareto distribution
with scale $v_{\mathrm{e}}$ and shape $\alpha_{\mathrm{e}}>2$,
\begin{equation}
P_{\mathrm{P},a}\left(\psi_\mathrm{e}\right)=\left\{ \begin{array}{c}
\alpha_{\mathrm{e}} \frac{v_{\mathrm{e}}^{\alpha_{\mathrm{e}}}}{\psi_{\mathrm{e}}^{\alpha_{\mathrm{e}}+1}},\\
0,
\end{array}\begin{array}{c}
\psi_\mathrm{e}\geq v_{\mathrm{e}}\\
\mathrm{otherwise}
\end{array}.\right.
\label{eq:P_P_psi_e_Pareto_1}
\end{equation}
The expected value of $\Psi_{\mathrm{e}}$ is $E_{\mathrm{P}}\left[\Psi_{\mathrm{e}}\right]=\frac{\alpha_{\mathrm{e}} v_{\mathrm{e}}}{\alpha_{\mathrm{e}}-1}$
b. Thus, if we set $v_{\mathrm{e}}=\frac{\alpha_{\mathrm{e}}-1}{\alpha_{\mathrm{e}}}r_a$, 
we obtain $E_{\mathrm{P}}\left[\Psi_{\mathrm{e}}\right]=r_{a}$ b, i.e., we
match the expected query volume per device to that of the Uniform distribution. The characterization of the energy consumption for queries with Pareto-distributed volumes is summarized in the following proposition.

\begin{prop}
\label{prop:Pareto_E}
The average energy consumption for Pareto-distributed media query volumes is given by
\begin{IEEEeqnarray}{rCl}
E_{\mathrm{exp,P}} & =&  \left[  g_{\mathrm{e}}  + i_{\mathrm{e}}  \left[  (\alpha_{\mathrm{e}}-1)^{\alpha_{\mathrm{e}}-1} c_{\mathrm{e}} (\alpha_{\mathrm{e}} c_{\mathrm{e}})^{-\alpha_{\mathrm{e}}} + c_{\mathrm{e}}-1  \right]    \right]r_{a}, \IEEEeqnarraynumspace
 \label{eq:E_exp_P_r_1}
\end{IEEEeqnarray}
and the one-sided variation of the energy consumption from idle mode to active mode is given by
\begin{IEEEeqnarray}{rCl}
E_{\mathrm{var,P}} & = & 2 g_{\mathrm{e}}^2 \frac{ (\alpha_{\mathrm{e}}-1)^{\alpha_{\mathrm{e}}-1}c_{\mathrm{e}}^{2 - \alpha_{\mathrm{e}}}  }{\alpha_{\mathrm{e}}^{\alpha_{\mathrm{e}}}(\alpha_{\mathrm{e}}-2)} r_{a}^{2}.
\label{eq:E_var_P_r_1}
\end{IEEEeqnarray}

\end{prop}

\begin{IEEEproof}
The expressions \eqref{eq:E_exp_P_r_1} and \eqref{eq:E_var_P_r_1} are obtained by substituting the Pareto PDF \eqref{eq:P_P_psi_e_Pareto_1} in \eqref{eq:E_exp} and \eqref{eq:E_var}, respectively, and deriving the closed-form result of the integral expressions. 
\end{IEEEproof}

Note that Proposition 2 assumes that $c_{\mathrm{e}} \geq \frac{\alpha_{\mathrm{e}}-1}{\alpha_{\mathrm{e}}}$, since, otherwise, the device will never switch from active to idle state. In this case, the optimal solution, $c_\mathrm{e,P,primary}$, of \eqref{eq:primal} cannot be expressed in closed form, but it is obtained via efficient convex optimization algorithms such as gradient descent.
On the other hand, from \eqref{eq:E_var_P_r_1}, we can derive the solution of \eqref{eq:dual} as
\begin{equation}
c_{\mathrm{e,P,dual}} = \left[  \frac{ \alpha_{\mathrm{e}}^{\alpha_{\mathrm{e}}} (\alpha_{\mathrm{e}} -2 )   E_\mathrm{max\_var}    }{   2 g_{\mathrm{e}}^2 (\alpha_{\mathrm{e}}-1) ^{\alpha_{\mathrm{e}}-1}   r_{a}^2  }  \right]^{1/(2-\alpha_{\mathrm{e}})}.
\label{eq:ce_opt_P}
\end{equation}


A particular case of interest for the Pareto distribution arises when  $\alpha_{\mathrm{e}} \to +\infty$: in this limit case, the query volume per device converges to the expectation $E_{\mathrm{P}}\left[\Psi_{\mathrm{e}}\right]=r_a$, i.e., to \textit{fixed-volume} query production per monitoring interval {and activity zone}. Then, since $c_{\mathrm{e}}\geq \frac{\alpha_{\mathrm{e}}-1}{\alpha_{\mathrm{e}}}$, as $\alpha_{\mathrm{e}} \to \infty$, the average energy consumption tends~to
\begin{equation}
E_{\mathrm{exp,P}} = \left[g_{\mathrm{e}} + i_{\mathrm{e}} (c_{\mathrm{e}}-1)\right]r_{a},
\end{equation}
 and the one-side energy variation from idle to active mode converges to zero (the device is in idle mode for a fixed part of every monitoring interval). Then, the activation threshold which meets the average energy consumption constraint $E_{\mathrm{max\_exp}}$ is given by
\begin{equation}
c_{\mathrm{e,P,primary}} = 1 + \frac{  E_{\mathrm{max\_exp}}  - g_{\mathrm{e}}r_{a}  }{i_{\mathrm{e}}  r_{a}},
\end{equation}
provided that $E_{\mathrm{max\_exp}} \geq g_{\mathrm{e}} r_{a}$ (which must hold or else the query production rate, $r_{a}$, is not achievable).


\subsubsection{Exponential distribution}

This distribution is relevant to our application context since the marginal statistics of compressed image and video traffic have often been modeled
as exponentially decaying \cite{daiMPEG4exponential}. Consider $P_{\mathrm{E},a}\left(\psi_\mathrm{e}\right)$
as the Exponential distribution with rate parameter ${1}/{r_{a}}$ {for each activity zone $a$}
\begin{equation}
P_{\mathrm{E},a}(\psi_{\mathrm{e}})  = \frac{1}{r_{a}} \exp \left(  - \frac{\psi_{\mathrm{e}}}{r_{a}}  \right),
\label{eq:P_P_psi_e_Exponential_1}
\end{equation}
for $\psi_{\mathrm{e}} \geq 0$. In this case, the expected value of $\Psi_{\mathrm{e}}$ is $E_{\mathrm{E}}\left[\Psi_{\mathrm{e}}\right]=r_{a}$ b. The characterization of the energy consumption for queries with exponentially distributed volumes is summarized in the following proposition.

\begin{prop}
\label{prop:Exponential_E}
The average energy consumption for Exponentially-distributed media query volumes is given by
\begin{equation}
E_{\mathrm{exp,E}} = \left[  g_{\mathrm{e}}  + i_{\mathrm{e}} \left(c_{\mathrm{e}} + e^{-c_{\mathrm{e}} } -1  \right)  \right] r_{a},
\label{eq:E_exp_exponential_1}
\end{equation}
and the one-sided variation of the energy consumption from idle mode to active mode is given by
\begin{equation}
E_{\mathrm{var,E}} = 2 g_{\mathrm{e}}^2 \mathrm{exp}(-c_{\mathrm{e}}) {r_{a}^2} .
\label{eq:E_var_exponential_1}
\end{equation}
\end{prop}
\begin{IEEEproof}
The expressions \eqref{eq:E_exp_exponential_1} and \eqref{eq:E_var_exponential_1} are obtained by substituting the Exponential PDF \eqref{eq:P_P_psi_e_Exponential_1} in \eqref{eq:E_exp} and \eqref{eq:E_var}, respectively, and deriving the closed-form result of the integral expressions. 
\end{IEEEproof}


In this case, the closed form solution of the problem \eqref{eq:primal} can be derived from \eqref{eq:E_exp_exponential_1} 
as
\begin{IEEEeqnarray}{rCl}
\nonumber
c_{\mathrm{e,E,primary}} & = & W_0 \left( - \exp\left(  -    (E_{\mathrm{max\_exp}} + i_{\mathrm{e}}r_{a} - g_{\mathrm{e}}r_{a}  )/(i_{\mathrm{e}}r_{a})   \right)      \right)\\
& & +   \left(E_{\mathrm{max\_exp}} + i_{\mathrm{e}}r_{a} - g_{\mathrm{e}}r_{a}  \right)/(i_{\mathrm{e}}r_{a}), \IEEEeqnarraynumspace
\end{IEEEeqnarray}
where $W_0(\cdot)$ is the main branch of the standard Lambert W function \cite{corless1996lambertw}. 
Analogously, from \eqref{eq:E_var_exponential_1} we can derive the closed form solution of \eqref{eq:dual} as
\begin{equation}
c_{\mathrm{e,E,dual}} =  \ln \frac{2 g^2_{\mathrm{e}}  r_{a}^2}{ E_\mathrm{max\_var} }.
\end{equation}

\subsubsection{Half-Gaussian distribution}

We consider now $P_{\mathrm{H},a}\left(\psi_{\mathrm{e}}\right)$
as the Half-Gaussian distribution with mean $E_{\mathrm{H}}\left[\Psi_{\mathrm{e}}\right]=r_{a}$ {for each activity zone $a$}
\begin{equation}
P_{\mathrm{H},a}\left(\psi_\mathrm{e}\right)=\left\{ \begin{array}{c}
\frac{2}{\pi r_{a}}  \exp \left(   -  \frac{\psi_{\mathrm{e}}^2}{\pi r_{a}^2}   \right)    ,\\
0,
\end{array}\begin{array}{c}
\psi_\mathrm{e}\geq 0\\
\mathrm{otherwise}
\end{array}.\right.\label{eq:P_P_psi_e_Half_1}
\end{equation}
This distribution has been widely used in data gathering problems
in science and engineering when the modeled data has non-negativity
constraints. Some recent examples include the statistical characterization
of motion vector data rates in Wyner-Ziv video coding algorithms suitable
for WSNs \cite{tagliasacchiWynerZivGaussian}, or the statistical
characterization of sample amplitudes captured by an image sensor
\cite{LamGoodmanDCT,andreopoulosincrementalHarris}.  The characterization of the energy consumption for queries with Half-Gaussian distributed volumes is summarized in the following proposition.

\begin{prop}
\label{prop:HalfGaussian_E}
The average energy consumption for Half-Gaussian-distributed media query volumes is given by
\begin{equation}
E_{\mathrm{exp,H}} = \left( g_{\mathrm{e}}  + i_{\mathrm{e}}  c_{\mathrm{e}} \mathrm{erf}\left(  \frac{c_{\mathrm{e}}}{ \sqrt{\pi} }   \right)   + i_{\mathrm{e}} \left( \exp\left(  - \frac{c_{\mathrm{e}}^2}{\pi }  \right) -1     \right)      \right) r_{a},
\label{eq:E_exp_H_r_1}
\end{equation}
where $\mathrm{erf}(\cdot)$ is the error function, and the one-side variation of the energy consumption from idle mode to active mode is given by
\begin{equation}
E_{\mathrm{var,H}}  =  \frac{g_{\mathrm{e}}^2 }{2} \left(   (2 c_{\mathrm{e}}^2 + \pi ) \left(1-  \mathrm{erf}\left(   \frac{c_{\mathrm{e}}}{ \sqrt{\pi} }  \right) \right) -2 c_{\mathrm{e}} \exp \left(     - \frac{c_{\mathrm{e}}^2}{\pi }  \right)   \right) r_{a}^2.
\label{eq:E_var_H_r_1}
\end{equation}
\end{prop}
\begin{IEEEproof}
The expressions \eqref{eq:E_exp_H_r_1} and \eqref{eq:E_var_H_r_1} are obtained by substituting the Half-Gaussian PDF \eqref{eq:P_P_psi_e_Half_1} in \eqref{eq:E_exp} and \eqref{eq:E_var}, respectively, and simplifying the integral expressions.
\end{IEEEproof}

In this case, the solutions to \eqref{eq:primal} and \eqref{eq:dual} cannot be expressed in closed form. However, they can be efficiently computed using gradient descent given that the error function can be efficiently and accurately approximated with well known methods~\cite{winitzki2003uniform}.


\subsubsection{Billing cost under Pareto, Exponential and Half-Gaussian distribution}
We now consider the billing cost for the processing of queries uploaded from $n$  devices via an IoT aggregator.
Let us first consider the aggregate query volume distribution modeled via {a Pareto distribution with mean $E_{\mathrm{P}}[\Psi_{\mathrm{b}}]=r_\mathrm{tot}$ [with $r_\mathrm{tot}$ given by \eqref{eq:r_tot}}], i.e., 
\begin{equation}
P_{\mathrm{P}}\left(\psi_\mathrm{b}\right)=\left\{ \begin{array}{c}
\alpha_{\mathrm{b}} \frac{v_{\mathrm{b}}^{\alpha_{\mathrm{b}}}}{\psi_{\mathrm{b}}^{\alpha_{\mathrm{b}}+1}},\\
0,
\end{array}\begin{array}{c}
\psi_\mathrm{b}\geq v_{\mathrm{b}}\\
\mathrm{otherwise}
\end{array}\right. ,
\label{eq:P_P_psi_b_Pareto_1}
\end{equation}
where $\alpha_{\mathrm{b}}>2$ {and $v_{\mathrm{b}}=\frac{\alpha_{\mathrm{b}}-1}{\alpha_{\mathrm{b}}}r _{\mathrm{tot}}.$}
\begin{prop}
\label{prop:Pareto_B}
The average billing cost incurred from processing Pareto-distributed query volumes is given by
\begin{equation}
B_{\mathrm{exp,P}}   =  
(g_{\mathrm{b}} - i_{\mathrm{b}}) r_\mathrm{tot} + (i_{\mathrm{b}} + p_{\mathrm{b}})  \frac{(\alpha_{\mathrm{b}}-1)^{\alpha_{\mathrm{b}}-1}}{\alpha_{\mathrm{b}}^{\alpha_{\mathrm{b}}}c_{\mathrm{b}}^{\alpha_{\mathrm{b}}-1}}r_{\mathrm{tot}}^{\alpha_{\mathrm{b}}}  + i_{\mathrm{b}} c_{\mathrm{b}}.
\end{equation}
The minimum billing cost is obtained when
\begin{equation}
c_{\mathrm{b,P}} =  
\left(   \frac{i_{\mathrm{b}}  + p_{\mathrm{b}}}{i_{\mathrm{b}}}   \right)^{\frac{1}{\alpha_{\mathrm{b}}}} \frac{\alpha_{\mathrm{b}}-1}{\alpha_{\mathrm{b}}} r_\mathrm{tot},
\label{eq:c_b,P_1}
\end{equation}
and it is given by
\begin{IEEEeqnarray}{rCl}
\min \{  B_{\mathrm{exp,P}}  \} & = & \left[ g_{\mathrm{b}} - i_{\mathrm{b}}   + i_{\mathrm{b}} \left(   \frac{i_{\mathrm{b}}+ p_{\mathrm{b}}}{i_{\mathrm{b}}}  \right)^{\frac{1}{\alpha_{\mathrm{b}}}}  \right] r_\mathrm{tot}.
 \label{eq:minBex}
\end{IEEEeqnarray}
\end{prop}

\begin{IEEEproof}
The proof stems from the evaluation of the general solution expressed in \eqref{eq:cb_opt_general} under the usage of the Pareto PDF.
\end{IEEEproof}

In order to ensure that the average billing cost is $B_{\mathrm{mean}}$ {when the maximum query volume constraint, $V_{\mathrm{max}}$, is satisfied, we first need to guarantee that
\begin{equation}
B_{\mathrm{mean}} \leq V_{\mathrm{max}} \left[ g_{\mathrm{b}} - i_{\mathrm{b}}   + i_{\mathrm{b}} \left(   \frac{i_{\mathrm{b}}+ p_{\mathrm{b}}}{i_{\mathrm{b}}}  \right)^{\frac{1}{\alpha_{\mathrm{b}}}}  \right].
\label{eq:feasibility_Pareto}
\end{equation}
Then, by determining the number of devices, $n_1,\ldots, n_A$, for activity zone based on proportional fairness, and by setting $\min \{  B_{\mathrm{exp,P}}\}=B_{\mathrm{mean}}$ in \eqref{eq:minBex}, we obtain
\begin{IEEEeqnarray}{rCl}
n_{\mathrm{P},a} & = & \frac{B_{\mathrm{mean}}}{\left[ g_{\mathrm{b}} - i_{\mathrm{b}}   + i_{\mathrm{b}} \left(   \frac{i_{\mathrm{b}}+ p_{\mathrm{b}}}{i_{\mathrm{b}}}  \right)^{\frac{1}{\alpha_{\mathrm{b}}}}  \right] A r_a},
\end{IEEEeqnarray}
where $B_{\mathrm{mean}}$ is bounded by the constraint of \eqref{eq:feasibility_Pareto}. If $B_{\mathrm{mean}}$ is set such that  $V_\mathrm{max}$ is expected to be fully utilized, i.e., \eqref{eq:feasibility_Pareto} becomes an equality, then $n_{\mathrm{P},a}$ is given by \eqref{eq:prop_fair}.}
We also note that, when assuming that the aggregate query volume is Pareto distributed, by letting $\alpha_{\mathrm{b} } \to +\infty$, we can analyze the case when the aggregate query volume at the IoT is fixed and {equal to $r_\mathrm{tot}$. In this case, if $c_{\mathrm{b}} \geq r_\mathrm{tot}$, the average billing cost is simply given by
\begin{equation}
B_{\mathrm{exp,P}} = (g_{\mathrm{b}}-i_{\mathrm{b}}) r_\mathrm{tot} + i_{\mathrm{b}} c_{\mathrm{b}},
\end{equation}
which is minimized by setting $c_{\mathrm{b}}$ equal to the mean, i.e., $c_{\mathrm{b,P}} = r_{\mathrm{tot}}$.}

Let us consider the aggregate query volume distribution modeled via an Exponential distribution with {mean $E_{\mathrm{E}}[\Psi_{\mathrm{b}}] = r_\mathrm{tot}$, i.e., 
\begin{equation}
P_{\mathrm{E}}(\psi_{\mathrm{b}})  = \frac{1}{r_\mathrm{tot}} \exp \left(  -\frac{1}{r_\mathrm{tot}} \psi_{\mathrm{b}}   \right),
\label{eq:P_P_psi_b_Exponential_1}
\end{equation}}
for $\psi_{\mathrm{b}} \geq 0$.
 \begin{prop}
 \label{prop:Exponential_B}
The average billing cost incurred from processing Exponentially-distributed query volumes is given by{
\begin{equation}
B_{\mathrm{exp,E}} = (g_{\mathrm{b}} - i_{\mathrm{b}}) r_\mathrm{tot} + i_{\mathrm{b}}c_{\mathrm{b}} +(i_{\mathrm{b}} + p_{\mathrm{b}}) r_\mathrm{tot} e^{-\frac{c_{\mathrm{b}}}{r_\mathrm{tot}}}.
\end{equation}
The minimum billing cost is obtained when
\begin{equation}
c_{\mathrm{b,E}} = {r_\mathrm{tot}}  \ln \frac{i_{\mathrm{b}} + p_{\mathrm{b}}}{i_{\mathrm{b}}},
\label{eq:c_b,E_1}
\end{equation}
and it is given by
\begin{equation}
\min \{ B_{\mathrm{exp,E}} \} = \left( g_{\mathrm{b}} + i_{\mathrm{b}} \ln \frac{i_{\mathrm{b}} + p_{\mathrm{b}}}{i_{\mathrm{b}}}    \right) r_\mathrm{tot}.
\label{eq:Bmin_E}
\end{equation}}
\end{prop}
\begin{IEEEproof}
The proof stems from the evaluation of the general solution expressed in \eqref{eq:cb_opt_general} under the usage of the Exponential PDF.
\end{IEEEproof}

In order to ensure that the average billing cost is $B_{\mathrm{mean}}$ {when the maximum query volume constraint $V_{\mathrm{max}}$ is satisfied, we first need to guarantee that
\begin{equation}
B_{\mathrm{mean}} \leq V_{\mathrm{max}} \left( g_{\mathrm{b}} + i_{\mathrm{b}} \ln \frac{i_{\mathrm{b}} + p_{\mathrm{b}}}{i_{\mathrm{b}}}    \right).
\label{eq:feasibility_Exponential}
\end{equation}
Then, by adopting proportional fairness to allocate the number of devices for each activity zone $n_1,\ldots, n_A$, and by setting $\min \{  B_{\mathrm{exp,P}}\}=B_{\mathrm{mean}}$ in \eqref{eq:Bmin_E}, we obtain
\begin{IEEEeqnarray}{rCl}
n_{\mathrm{E},a} & = & \frac{B_{\mathrm{mean}}}{   \left( g_{\mathrm{b}} + i_{\mathrm{b}} \ln \frac{i_{\mathrm{b}} + p_{\mathrm{b}}}{i_{\mathrm{b}}}    \right)  A r_a},
\end{IEEEeqnarray}
where the value of $B_{\mathrm{mean}}$ is upper-bounded by \eqref{eq:feasibility_Exponential}. Similarly as before, if \eqref{eq:feasibility_Exponential} is met with equality, then $n_{\mathrm{E},a}$ is given by the simple solution of \eqref{eq:prop_fair}.}

Finally, consider the case when the aggregate query volume is  Half-Gaussian distributed with {mean $E_{\mathrm{H}} [\Psi_{\mathrm{b}}] = r_\mathrm{tot}$, i.e., 
\begin{equation}
P_{\mathrm{H}}\left(\psi_\mathrm{b}\right)=\left\{ \begin{array}{c}
\frac{2}{\pi r_\mathrm{tot}}  \exp \left(   -  \frac{\psi_{\mathrm{b}}^2}{\pi r_\mathrm{tot}^2}   \right)    ,\\
0,
\end{array}\begin{array}{c}
\psi_\mathrm{b}\geq 0\\
\mathrm{otherwise}
\end{array}.\right.
\label{eq:P_P_psi_b_Half_1}
\end{equation}}

\begin{prop}
\label{prop:HalfGaussian_B}
The average billing cost incurred from processing Half-Gaussian-distributed query volumes is {given by
\begin{IEEEeqnarray}{rCl}
\nonumber
B_{\mathrm{exp,H}} & = & (g_{\mathrm{b}} +  p_{\mathrm{b}})r_\mathrm{tot} - p_{\mathrm{b}} c_{\mathrm{b}} + (i_{\mathrm{b}} + p_{\mathrm{b}}) \\
& &  \times \left(    c_{\mathrm{b}} \mathrm{erf}\left(  \frac{c_{\mathrm{b}}}{ \sqrt{\pi} r_\mathrm{tot}}   \right)   + r n \left( \exp\left(  - \frac{c_{\mathrm{b}}^2}{\pi r_\mathrm{tot}^2}  \right) -1     \right)   \right). \IEEEeqnarraynumspace
\end{IEEEeqnarray}
The minimum billing cost is obtained when
\begin{equation}
c_{\mathrm{b,H}} = r_\mathrm{tot} \sqrt{\pi}  \mathrm{erf}^{-1}\left( \frac{p_\mathrm{b}}{p_{\mathrm{b}}+ i_{\mathrm{b}}}  \right),
\label{eq:c_b,H_1}
\end{equation}
and it is given by
\begin{IEEEeqnarray}{rCl}
\nonumber
\min \{ B_{\mathrm{exp,H}} \} & = & r_\mathrm{tot} \left[ g_{\mathrm{b}} -i_{\mathrm{b}} + (i_{\mathrm{b}} + p_{\mathrm{b}}) \right. \\
& & \left. \times   \exp \left(  - \left(  \mathrm{erf}^{-1} \left(  \frac{p_\mathrm{b}}{p_{\mathrm{b}}+ i_{\mathrm{b}}}   \right)    \right)^2  \right)        \right] .
\label{eq:Bmin_H}
\end{IEEEeqnarray}}
\end{prop}
\begin{IEEEproof}
The proof stems from the evaluation of the general solution expressed in \eqref{eq:cb_opt_general} under the usage of the Half-Gaussian PDF.
\end{IEEEproof}

In order to ensure that the average billing cost is $B_{\mathrm{mean}}$ {when the maximum query volume constraint $V_{\mathrm{max}}$ is satisfied, we first need to guarantee that
\begin{IEEEeqnarray}{rCl}
\nonumber
B_{\mathrm{mean}} &\leq& V_{\mathrm{max}} \left[ g_{\mathrm{b}} -i_{\mathrm{b}} + (i_{\mathrm{b}} + p_{\mathrm{b}}) \right. \\
& & \left. \times   \exp \left(  - \left(  \mathrm{erf}^{-1} \left(  \frac{p_\mathrm{b}}{p_{\mathrm{b}}+ i_{\mathrm{b}}}   \right)    \right)^2  \right)        \right] .
\label{eq:feasibility_HalfGaussian}
\end{IEEEeqnarray}
Via a proportionally-fair allocation of the number of devices for each activity zone [and by setting $\min \{  B_{\mathrm{exp,P}}\}=B_{\mathrm{mean}}$ in \eqref{eq:Bmin_H}], we obtain
\begin{equation}
n_{\mathrm{H},a}  = \frac{  B_{\mathrm{mean}}  }{ \left[ g_{\mathrm{b}} -i_{\mathrm{b}} + (i_{\mathrm{b}} + p_{\mathrm{b}}) \exp \left(  - \left(  \mathrm{erf}^{-1} \left(  \frac{p_\mathrm{b}}{p_{\mathrm{b}}+ i_{\mathrm{b}}}   \right)    \right)^2  \right)        \right]  A r_a },
\end{equation}
where the value of $B_{\mathrm{mean}}$ is upper-bounded by \eqref{eq:feasibility_HalfGaussian}. If the billing is set such that $V_\mathrm{max}$ is expected to be fully utilized, i.e., \eqref{eq:feasibility_HalfGaussian} is met with equality, then $n_{\mathrm{H},a}$ is given by \eqref{eq:prop_fair}, i.e., mutual coupling is achieved between the energy consumption parameters ($E_\mathrm{exp,U}$ and $E_\mathrm{var,U}$), the desired cloud billing cost ($B_\mathrm{mean}$), and the aggregator's data transmission volume ($V_\mathrm{max}$).}
%

{
\subsection{Discussion\label{sub:discussion}}
The results of this section can be used in practical applications to assess the impact in the required energy rates when the statistics of the query generation and transmission follow a certain PDF and the cloud billing costs are fixed. Conversely, if a particular IoT device technology is chosen, under the knowledge of the system and data gathering parameters, one can establish the appropriate cloud billing rates and the number of devices to include in each IoT aggregator in order to minimize the cloud infrastructure cost. Finally, for given IoT and cloud infrastructure parameters, one can assess the achievable query generation and transmission rates such that the IoT cluster leads to the optimal coupling between energy consumption and cloud billing cost. 
Thus, as shown in Fig. \ref{fig:framework-diagram}, our analytic results allow for the linkage of energy, data gathering and cloud billing parameters within IoT clusters of devices. Hence, our analysis can be used for early-stage
exploration of the capabilities of a particular IoT infrastructure, in conjunction with the data gathering requirements of a particular application, prior to embarking in cumbersome development and testing in the field.

\begin{figure}
\begin{centering}
\includegraphics[scale=0.19]{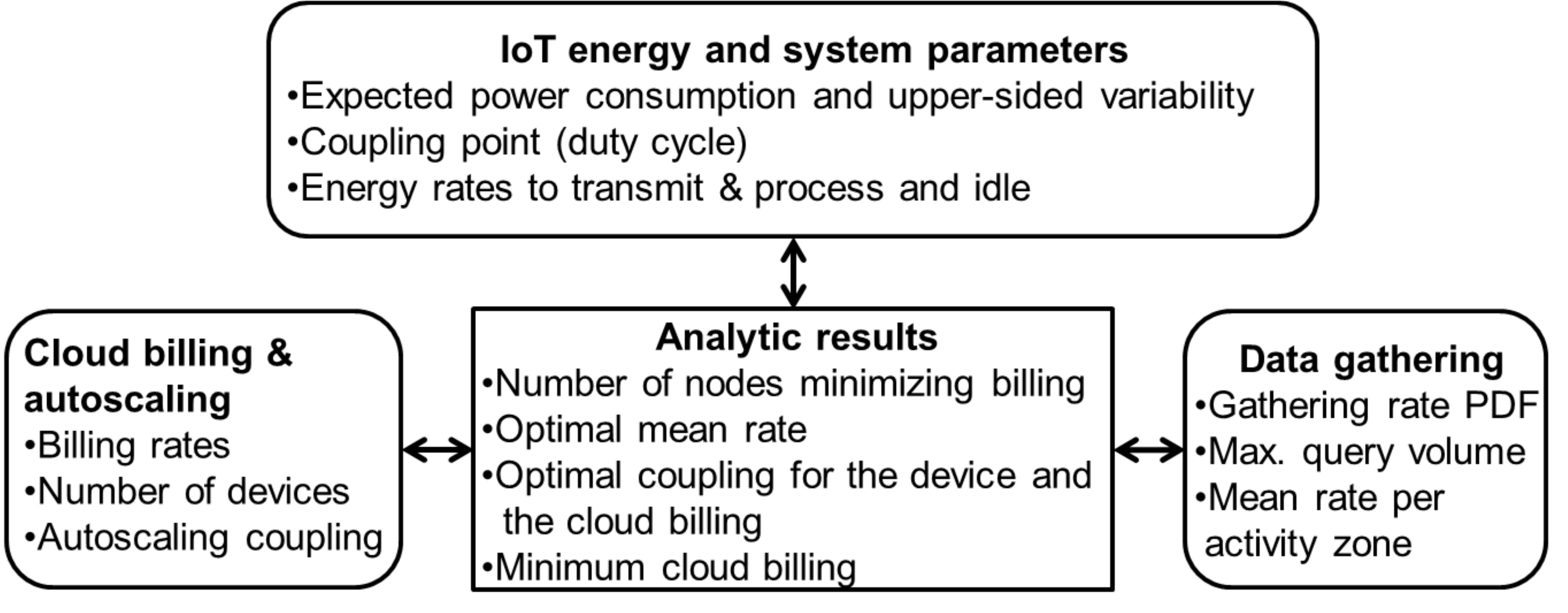} 
\par\end{centering}

\caption{Conceptual illustration of the linkage between: IoT system parameters, cloud billing \&\ autoscaling, and data gathering. When parameters from two out of three domains are provided, our analytic framework can be used to tune the parameters of the third. 
  \label{fig:framework-diagram}}
\end{figure}

}

\section{Evaluation of the Analytic Results\label{sec:Applications}}

To validate the proposed analytic modeling framework of
Propositions \ref{prop:Pareto_E}--\ref{prop:HalfGaussian_B}, we performed a series of experiments based on a visual
sensor network connected to an IoT aggregator, and eventually to an AWS S3 repository plus EC2 cluster of on-demand instances. The following subsections present the hardware and application specifications, as well as the achieved results.  

\subsection{System Specification}
We utilized a visual sensor network composed of multiple BeagleBone Linux embedded platforms \cite{redondi2014energy,CancliniSENSYS2013}. Each
BeagleBone is equipped with a RadiumBoard CameraCape board to provide for
the video frame acquisition. For energy-efficient processing, we downsampled
all input images to QVGA ($320 \times 240$) resolution. Further, our deployment involved: 
\begin{enumerate} 
\item
a portable computer acting as the IoT aggregator, i.e., collecting all bitstreams via a star topology with {$n_{\mathrm{tot}}=12$} nodes and the recently-proposed (and available as
open source) TFDMA protocol \cite{burana2012DTFDMA} for contention-free
MAC-layer coordination; 
\item an AWS\ S3 bucket where the IoT aggregator continuously uploads all queries via a TCP/IP connection using a Cron Job and the AWS\ Command Line Interface; 
\item
one reserved AWS instance running as the control server and assigning query volumes from S3 to AWS\ EC2 on-demand instances that serve as compute units; via AWS\ Auto Scaling, within each monitoring instance of $T$\ seconds, the number of on-demand instances is set to: 
\begin{itemize}
\item
3 when the query volume is below $c_\mathrm{b}$ b (``idle'' case). 
\item
30 when the volume exceeds $c_\mathrm{b}$ b (``active'' case).  
\end{itemize}
Under our deployment and the utilized application, the uploaded query vectors are matched with the feature vectors extracted from $80,000$ images of similar content. The corresponding billing rates per query bit for this matching operation were found to be  $i_\mathrm{b}=6.27\times 10^{-11}$ \$/b and $p_\mathrm{b}=6.27\times 10^{-10}$ \$/b. Regarding query traffic upload and storage costs, the corresponding billing rate per query bit was found to be  $g_\mathrm{b}=2.09\times 10^{-10}$ \$/b.  
\end{enumerate}
We note that no WiFi or other IEEE802.15.4
networks were concurrently operating in the utilized channels of the 2.4
GHz band. However, even if IEEE 802.11 or other IEEE 802.15.4 networks coexist
with the proposed deployment, well-known channel hopping schemes like
TSCH \cite{TSCH} can be used at the MAC layer to mitigate such external interference. Moreover, experiments have shown that such protocols
can scale to hundreds or even thousands of nodes \cite{TSCH}.
Therefore, our evaluation is pertinent to such scenarios that may
be deployed in the next few years within an IoT paradigm \cite{gubbi2013IoT}.
\subsection{Visual Similarity Identification Based on the Vector of Locally Aggregated Descriptors (VLAD) }
Each BeagleBone runs a basic motion detection algorithm (based on successive frame differencing) that generates a visual query only when sufficient motion is detected between  the captured video frames. The query vectors were generated using the state-of-the-art VLAD algorithm of Jegou \textit{et. al.} \cite{jegou2012aggregating}, which is based on SIFT feature extraction and compaction using local feature centers and a PCA projection matrix, both of which are derived offline via training with representative video data \cite{jegou2012aggregating}. The VLAD descriptor (i.e., query) size was set to 8192 b (256 coefficients of 32 b each).  

With respect to the visual feature extraction, dedicated energy-measurement tests were performed with the Beaglebone following the energy measurement setup of our previous work  \cite{redondi2014energy} (repeated tests with a resistor in series to the Beaglebone board and a high-frequency oscilloscope to capture the power consumption profile across repeated monitoring intervals). Under the utilized setup, we measured the average energy cost to produce and transmit a query bit, as well as the average
initialization cost per frame for both application scenarios. The resulting energy rates were: $g_\mathrm{e}=1.78\times 10^{-6}$ J/b and $i_\mathrm{e}=6.10\times 10^{-7}$ J/b. Moreover, under the utilized application, the Beaglebone can generate up to 1 query per second while being constantly active, i.e., $8192T$ b per monitoring interval of $T$ seconds. By setting mean query rates such that $E[\Psi_\mathrm{e}]\leq2048 T$ b  (i.e., up to $0.25$ queries per second), this allows for    $c_\mathrm{e}\in(0,4)$. In practice, we  restricted the utilized values for $c_\mathrm{e}$ to $(0,2]$ since higher values lead to the frame acquisition frequently exceeding 1 frame per second, which can lead to system instability.

\subsection{Results with Controlled Query Generation that Matches the Marginal PDFs Considered in the Theoretical Analysis}

\label{subsec:model_validation}

Under the settings described previously, our first goal is to validate the analytic expressions of Section \ref{sec:Characterization-of-Energy} that form the mathematical foundation
for Propositions \ref{prop:Pareto_E}--\ref{prop:HalfGaussian_E}.
To this end, we create a controlled query data production process
on each node by: \emph{(i)} artificially setting several sets
of query volumes according to the marginal PDFs of Section \ref{sec:Characterization-of-Energy}
via rejection sampling \cite{gilks1992adaptive}, a.k.a., Monte Carlo sampling; \emph{(ii)} setting
the mean query volume size per monitoring interval, $r$, to predetermined values. The
sets containing query volume sizes are preloaded onto the memory of
each sensor node during the setup phase. At run time, each BeagleBone node runs a special routine, which, per monitoring interval $t$: \textit{(i)} reads
the corresponding query volume size, $v(t)$, from the preloaded set; \textit{(ii)} captures and processes  $\frac{v(t)}{8192}$ frames, \textit{(iii)} transmits the produced $v(t)$ query bits to the IoT aggregator; \textit{(iv)} if $v(t)<c_\mathrm{e}E[\Psi_\mathrm{e}]$, captures and processes    $\frac{c_\mathrm{e}E[\Psi_\mathrm{e}]-v(t)}{8192}$ additional frames without transmitting queries. In this way, we emulate the actual operation of the node under various query volumes that match the statistical models considered by our analysis, and various thresholds $c_\mathrm{e}$ for switching between ``idle'' and ``active'' states. This controlled experiment is designed to confirm the validity of our analytic derivations when the operating conditions match the model assumptions precisely. 

\begin{figure}
\begin{centering}
\includegraphics[scale=0.48]{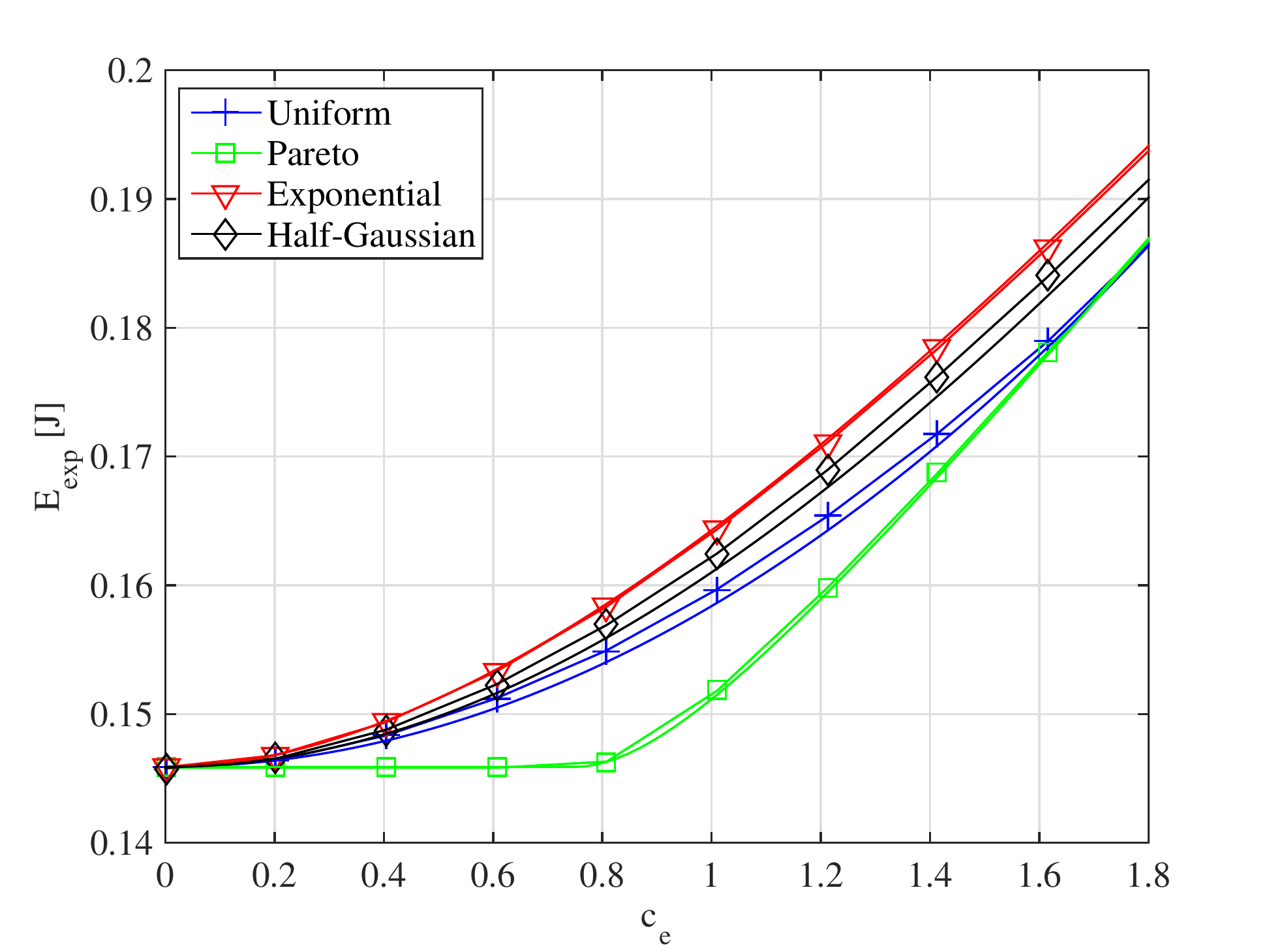} 
\par\end{centering}
\caption{Average energy consumption $E_{\mathrm{exp}}$ vs. $c_{\mathrm{e}}$. The average query volume was set to $r= 81,920$ b. For the case of Pareto distribution, we used $\alpha_{\mathrm{e}}=4$. Lines with markers: Monte Carlo experiments; Lines without markers: theoretical predictions. 
\label{fig:Eexp_MC}}
\end{figure}

\begin{figure}
\begin{centering}
\includegraphics[scale=0.48]{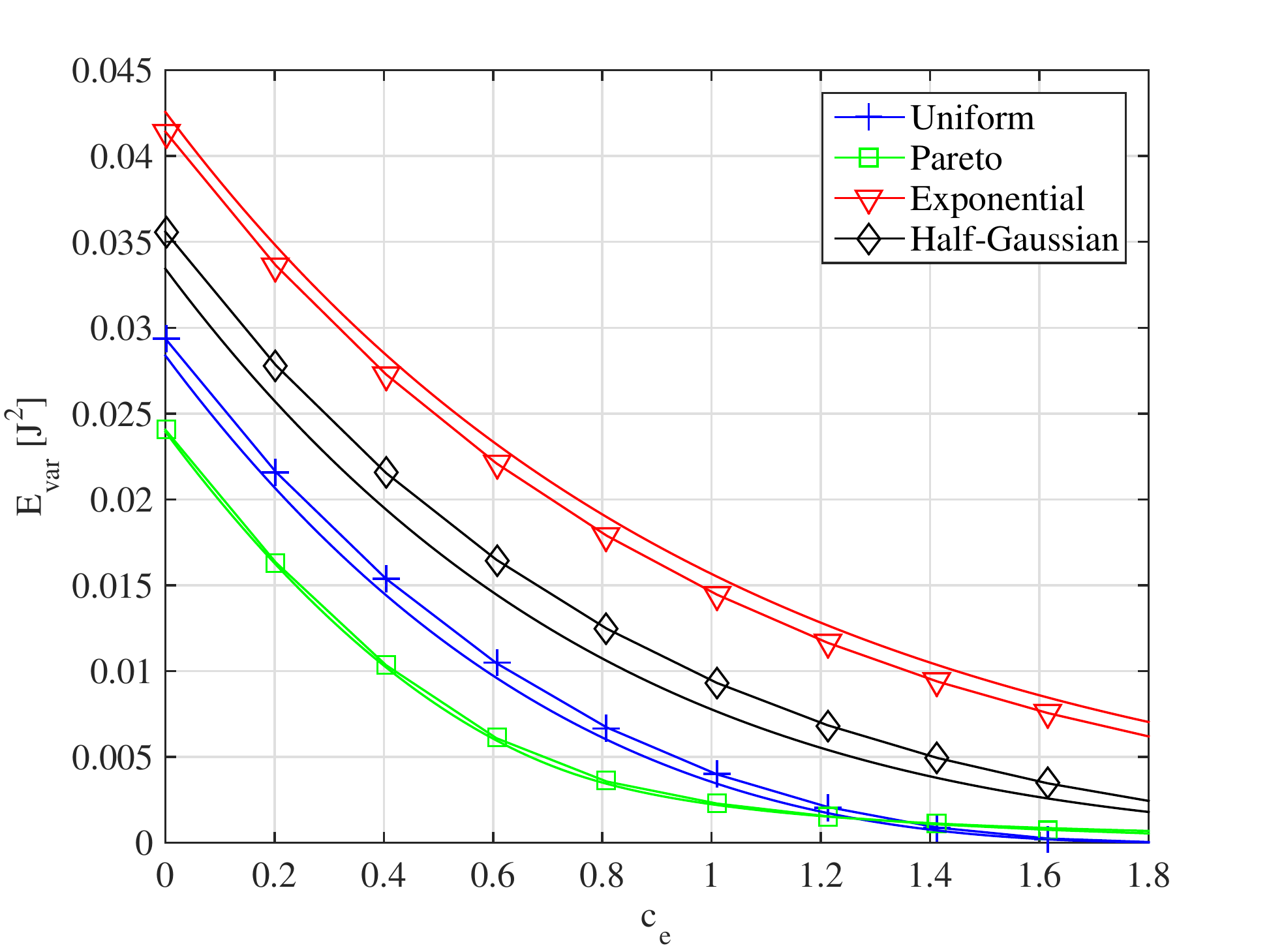} 
\par\end{centering}
\caption{One-sided energy consumption $E_{\mathrm{var}}$ vs. $c_{\mathrm{e}}$. The average query volume was set to $r= 81,920$ b. For the case of Pareto distribution, we used $\alpha_{\mathrm{e}}=4$. Lines with markers: Monte Carlo experiments; Lines without markers: theoretical predictions.
\label{fig:Evar_MC}}
\end{figure}

Indicative experimental results for monitoring time interval of $T=60$ s are reported in  Fig. \ref{fig:Eexp_MC} and Fig. \ref{fig:Evar_MC} for $r=81,920$ b. It is evident that
the theoretical results match the Monte Carlo experiments regarding energy consumption for all the
tested distributions, with all the  $R^{2}$
values (coefficients of determination) between the experimental and the model points being above $0.9964$.
We have observed the same level of accuracy
for the proposed model under a variety of data sizes ($r$) and active
time interval durations ($T$), but
omit these repetitive experiments for brevity of exposition.

Similar experiments have been carried out in order to validate the analytic expressions of Propositions  \ref{prop:Pareto_B}--\ref{prop:HalfGaussian_B} regarding the average billing cost. Specifically, we have submitted indicative queries to the cloud-computing service with volumes that have been generated according to the marginal PDFs of Section \ref{sec:Characterization-of-Energy}
via rejection sampling under various numbers of devices per IoT cluster ({$n_{\mathrm{tot}}$}) and various average query volumes. The aggregated queries are uploaded to the dedicated S3 bucket for the service and are processed by a number of instances that is controlled by the AWS Auto Scaling rules stated in the previous subsection. In this case, we used $T=600$ s and varied the value of $c_\mathrm{b}$ in order to see the incurred infrastructure billing costs under a variety of Auto Scaling thresholds.

\begin{figure}
\begin{centering}
\includegraphics[scale=0.48]{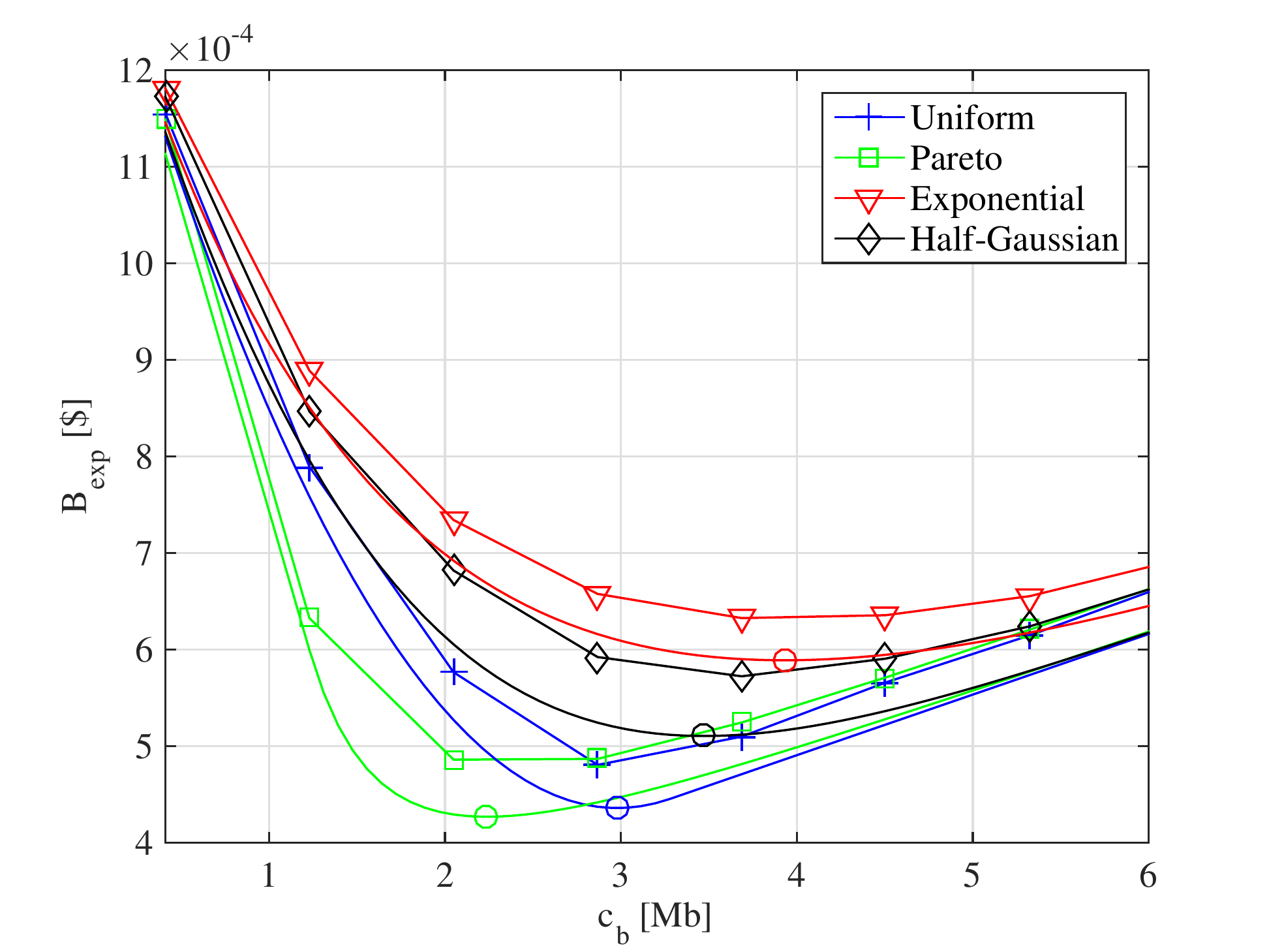} 
\par\end{centering}
\caption{Average billing cost $B_{\mathrm{exp}}$ vs. $c_{\mathrm{b}}$. The average query volume per device was set to $r= 163,840$ b and the experiment corresponds to $n=10$ devices. For the case of Pareto distribution, we used $\alpha_{\mathrm{e}}=4$. Lines with markers: Monte Carlo experiments; Lines without markers: theoretical predictions. The circles indicate minimum billing values as predicted by the analysis in Section \ref{sec:Characterization-of-Energy}.
 \label{fig:Bexp_MC}}
\end{figure}

Fig.~\ref{fig:Bexp_MC} presents indicative results under this setup. Evidently, the theoretical results follow the trends of the experimental data, with  $R^2$ coefficients being above $0.9947$ for all the distributions under consideration. However, the theoretical predictions tend to always underestimate the experimental values. This underestimation is due to the fact that our analysis does not take into account some practical latency and cost aspects of the service, for example that switching between ``idle'', ``active'' states is not instantaneous and other cost overheads (such as the cost of the control server) are not taken into account by our analysis. Similar results to Fig. \ref{fig:Bexp_MC} have been obtained for a variety of average query volumes and monitoring intervals, but are omitted for brevity of exposition.

\subsection{Results with {Real} Data}
\label{par:UserData}

We now present results when repeating the visual query generation, transmission and cloud-based processing for 25 monitoring intervals under a practical deployment within several research staff offices of the Electronic and Electrical Engineering Department of University College London. 
{The deployment environment comprises a large shared office space, which is composed of areas with low query generation activity (seated desk areas with low movement of people) and areas with high query generation activity (corridor areas with high movement of people).}
\subsubsection{Accuracy of energy estimation under a fixed setup {and one activity zone}}
 In the first batch of tests, each device's camera is set to fixed capture rate of 5 frames per second. Via successive frame differencing for motion detection, VLAD queries where generated when the contents of frames varied beyond a preset threshold, e.g., when people passed (or moved) in front of the device camera.  Back-end query similarity identification was done using prestored VLAD signatures of $80,000$ images of similar content based on the AWS setup described in the previous subsection.  

\begin{figure}
\begin{centering}
\includegraphics[scale=0.48]{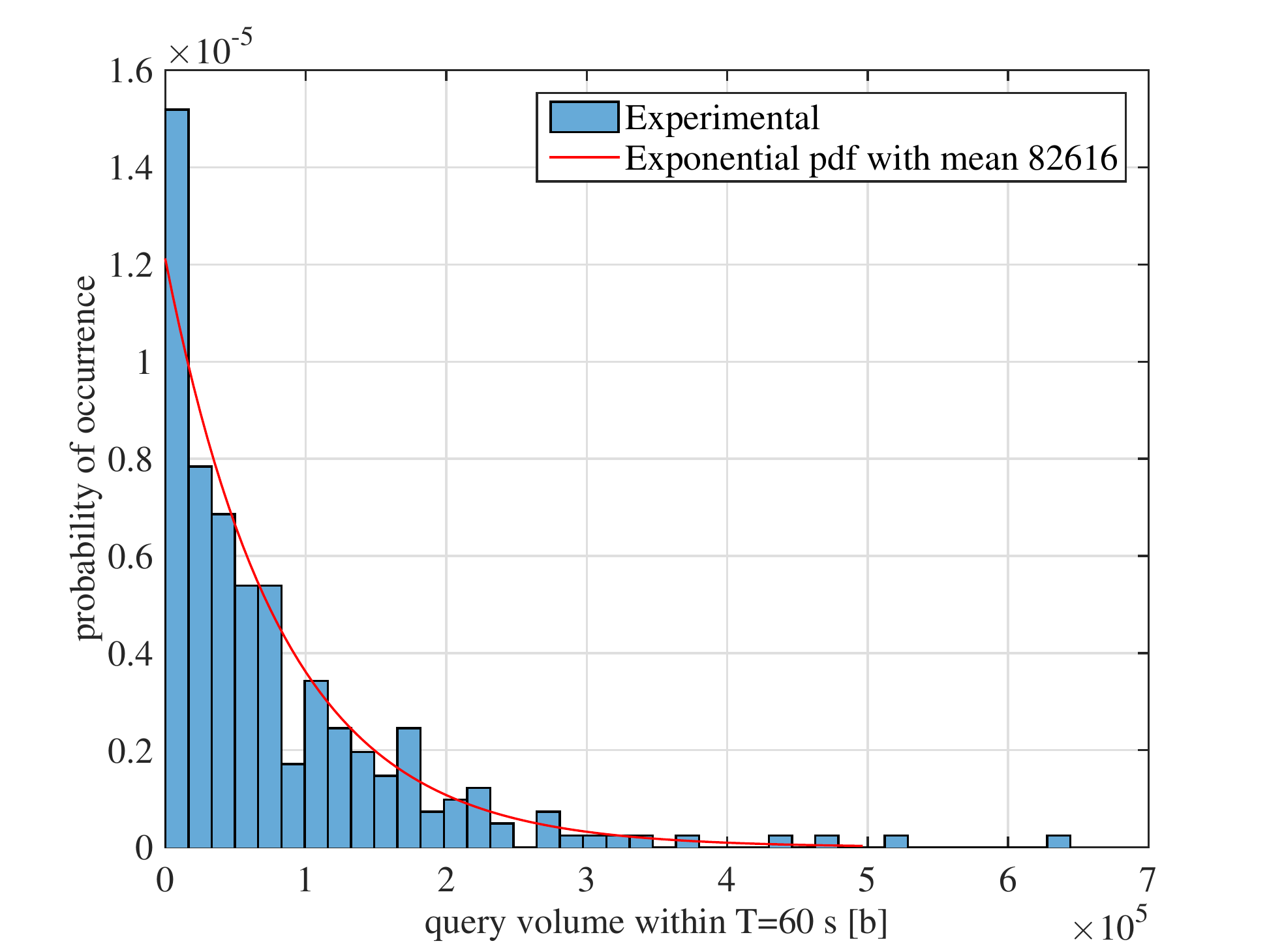} 
\par\end{centering}
\caption{Probability histogram of query volume for $T=60$ s {generated from our deployment experiments} and the best fit obtained via the Exponential distribution.  \label{fig:fit-exponential-pdf}}
\end{figure}

Once data has been collected, we fitted%
\footnote{Fitting is performed by matching the average data size $r$ of each
distribution to the average data size of the JPEG compressed frames
or the set of visual features.%
} the query production volumes to one of the distributions used
in Section \ref{sec:Characterization-of-Energy}, {i.e., assuming only one activity zone}. In the performed experiment, and under monitoring interval of $T=60$ s for the devices, we found
that the {real data} query volume histogram agreed best with the Exponential distribution with  $r=82,616$ b. For $T=\{600,1200\}$ s, the best fit was found to be the Pareto distribution with: $r=816,250$ b and $\alpha=3.89$, and  $r=1,569,700$ b and $\alpha=3.95$, respectively. An example for the fit obtained with the Exponential distribution is given in Fig. \ref{fig:fit-exponential-pdf}. Moreover, with respect to $c_\mathrm{e}$, we found that, under the acquisition of 5 frames per second, the system switched between ``idle'' and ``active'' states at $c_\mathrm{e}\cong 0.75$.

Under this setup and with the fitted values for Exponential and Pareto PDFs, Table \ref{tab:energy-results} presents the obtained experimental and theoretical values (via Propositions \ref{prop:Pareto_E} and \ref{prop:Exponential_E})\ for the expected energy and the one-sided energy variance for two monitoring intervals. It is observed that, despite the modeling mismatch due to the fitting shown in Fig. \ref{fig:fit-exponential-pdf}, the theoretical predictions on the expected energy consumption are always within 2\% of\ the experimentally-derived values, whereas theoretical predictions on the one-sided energy variation are within 20\% of experimental data. As such, the proposed energy-consumption model
can be used for early-stage testing of feasible application
deployments with respect to their energy consumption in order to determine
the impact of various options, prior
to time-consuming experimentation in the field. 

\subsubsection{Energy and billing cost reduction under parameter tuning {and two activity zones}}

We consider a further example to showcase how the theoretical modeling presented in Section \ref{sec:Characterization-of-Energy} can be leveraged in order to allow parameter tuning within  $T=\{600,1200\}$. When $T=600$ s, we set the maximum admissible one-sided energy variation to $E_{\mathrm{max\_var}}=0.35$ J$^2$, whereas, when $T=1200$ s, we set $E_{\mathrm{max\_var}}=1$ J$^2$. We report experimental results {generated from our deployment} for a scenario where $A=2$ different zones are present, corresponding to ``low'' and ``high'' activity. This was achieved by positioning some devices in areas with low movement of people (seated desk area of a large shared office)\ and some others in areas with high movement (corridor area of a large shared office). The devices contained in the first (i.e., ``low activity'') zone produce query volumes that are best approximated by the Pareto distribution with $r_1=160,000$ b and $\alpha_1=2.42$, if $T=600$ s, and  $r_1=320,000$ b and $\alpha_1=2.58$, if $T=1200$ s. On the other hand, the devices in the second (i.e., ``high activity'') zone are best approximated via the Pareto distribution, albeit with $r_2=4,915,600$ b and $\alpha_2=3.27$, if $T=600$ s, and  $r_2=9,572,600$ b and $\alpha_2=4.10$, if $T=1200$ s. 
The IoT aggregator admits $n_1=10$ devices from the low-activity zone and $n_2=2$ devices from the high-activity zone, thus resulting in a total query volume occupation of $11.43$ Mb, when $T=600$ s, and $22.35$ Mb, when $T=1200$ s. Instead of presetting the frame acquisition to fixed value (5 frames per second, which led to $c_\mathrm{e}\cong 0.75$), we now change the acquisition rate, thereby controlling the activation threshold $c_\mathrm{e}$. Our aim is to set $c_{\mathrm{e}}$ to the value that minimizes the expected energy consumption while verifying the constraint on the one-sided energy variation, which is given by \eqref{eq:ce_opt_P}. We then compare the expected energy consumption obtained with such setting, with the one obtained via two {baseline} solutions that impose $c_{\mathrm{e}}=1.5$ or $c_{\mathrm{e}}=2$ (corresponding to acquiring 10 and 13.3 frames per second). The obtained values from \eqref{eq:ce_opt_P} were found to be $c_{\mathrm{e,P,dual}}=0.82$, corresponding to capturing 5.5 frames per second, for the case $T=600$ s, and $c_{\mathrm{e,P,dual}}=0.92$, corresponding to capturing 6 frames per second, for the case $T=1200$ s. The obtained energy consumption results, reported in Table~\ref{tab:energy-data-results}, show that by selecting  $c_{\mathrm{e}}$ via the proposed analytic framework, we can achieve gains of up to $23,55 \%$ with respect to baseline settings. {It is important to note that, beyond the presented comparisons of Table~\ref{tab:energy-data-results}, the optimal tuning of $c_\mathrm{e}$ always led to decreased energy consumption in comparison to all other baseline settings attempted, thereby experimentally confirming the validity of Propositions 1 and 2.} 

Let us now consider the billing cost associated to the cloud infrastructure.  
%
Under the utilized setup, we determined the autoscaling threshold, $c_\mathrm{b}$, that is expected to lead to the minimum cloud infrastructure billing cost based on Proposition~\ref{prop:Pareto_B}. We then benchmarked the obtained cost of the system under this  threshold against the intuitive (albeit \textit{ad-hoc}) {baseline} setting of $c_\mathrm{b}=r_{\mathrm{tot}}=r_1 n_1 + r_2 n_2$, which corresponds to the autoscaling threshold being set to match the average query volume of all $n_{\mathrm{tot}}$ devices. The results, given in Table \ref{table:B}, show that the obtained billing cost is {$14\%$ (for $T=600$ s) and $12\%$ (for $T=1200$ s) lower} than the case of the same query volume processing under the baseline autoscaling threshold. This demonstrates that establishing the system parameters based on the theoretical analysis can lead to important cost savings within cloud-based media query processing systems. {Importantly, the optimal values derived by \eqref{eq:c_b,P_1}  have consistently outperformed all other baseline settings attempted, thereby experimentally confirming the validity of Proposition 5.}\

\begin{table}
\caption{\label{tab:energy-results}Expected energy consumption and one-sided variation. Experimental results {from our deployment} and theoretical prediction, $c_{\mathrm{e}}=0.75$.}
\begin{center}
\begin{tabular}{r|m{22mm}m{22mm}}
 &Theoretical& Experimental \\
 \toprule
  $T=60$ s  & $E_{\mathrm{exp}} = 0.1588 $ J $E_{\mathrm{var}} = 0.0201$ J$^2$& $E_{\mathrm{exp}} = 0.1603$ J $E_{\mathrm{var}} = 0.0254$ J$^2$ \\
 \midrule
$T = 1200$ s &   $E_{\mathrm{exp}} = 2.7965 $ J $E_{\mathrm{var}} = 1.4349$ J$^2$ & $E_{\mathrm{exp}} = 2.8440$ J $E_{\mathrm{var}} = 1.2234$ J$^2$ \\
\bottomrule
\end{tabular}
\label{table:E}
\end{center}
\end{table}

\begin{table}
\caption{\label{tab:energy-data-results}Expected energy consumption with one-sided variation constraint generated from our deployment experiments. The {baseline} solutions correspond to setting $c_{\mathrm{e}}$ equal to $1.5$ or $2$. The proposed solution is obtained with $c_{\mathrm{e}}=0.82$ and $c_{\mathrm{e}}=0.92$, derived via \eqref{eq:ce_opt_P}.}
\begin{center}
\begin{tabular}{m{20mm}|m{16.5mm}m{16.5mm}|m{16.5mm}}
  & $c_{\mathrm{e}} =1.5$ & $c_{\mathrm{e}} = 2$& $c_{\mathrm{e}}$ as in \eqref{eq:ce_opt_P}  \\
 \toprule
 $T=600$\,s $E_{\mathrm{var}}\leq0.35$\,J$^2$   &    $E_{\mathrm{exp}} = 1.72$\,J $(13.26 \%)$  & $E_{\mathrm{exp}} = 1.95  $\,J $(23,55 \%)$  & $E_{\mathrm{exp}} = 1.49$\,J ($c_{\mathrm{e}}=0.82$)   \\
 \midrule
$T = 1200$\,s  $E_{\mathrm{var}}\leq1.00$\,J$^2$& $E_{\mathrm{exp}} = 3.30$\,J $(12,40 \%)$  & $E_{\mathrm{exp}} = 3.75  $\,J $(22,95 \%)$  & $E_{\mathrm{exp}} = 2.89  $\,J ($c_{\mathrm{e}}=0.92$) \\
\bottomrule
\end{tabular}
\label{table:Edata}
\end{center}
\end{table}

\begin{table}
\caption{\label{tab:billing-results}Expected billing cost generated from our deployment experiments. The baseline solution corresponds to setting $c_{\mathrm{b}} =r_1 n_1 + r_2 n_2$. The proposed solution is obtained with $c_{\mathrm{b}}$ as in Proposition~\ref{prop:Pareto_B}.}
\begin{center}
{
\begin{tabular}{m{14mm}|m{25mm}m{25mm}|m{8mm}}
  & Baseline & Proposition~\ref{prop:Pareto_B} & Saving \\
 \toprule
 $T=600$~s    $n_1=10~~~~~$ $n_2=2$   &    $B_{\mathrm{exp}} = 3.38 \cdot 10^{-3}$ \$ $c_\mathrm{b}=11.43$ Mb & $B_{\mathrm{exp}} = 2.89 \cdot 10^{-3}  $ \$ $c_\mathrm{b}=14.90$ Mb  & 14 \% \\
 \midrule
$T = 1200$~s $n_1=10~~~~~$ $n_2=2$   & $B_{\mathrm{exp}} = 5.86 \cdot 10^{-3} $ \$  $c_\mathrm{b}=22.35$ Mb& $ B_{\mathrm{exp}} =5.15 \cdot 10^{-3}  $ \$  $c_\mathrm{b}=27.75$ Mb & 12 \%  \\
\bottomrule
\end{tabular}
\label{table:B}
}
\end{center}
\end{table}

\section{Conclusions\label{sec:Conclusions}}
We propose a novel theoretical framework for establishing trade-offs in the energy consumption and infrastructure billing cost of Internet-of-Things\ oriented deployments comprising mobile devices generating media queries that are processed by a back-end cloud computing service. Our analysis incorporates energy consumption and cloud infrastructure billing rates when the devices and the cloud computing system adapt their resource consumption according to the volume of generated queries by switching between ``idle'' and ``active'' states.  Experiments with an embedded platform and Amazon Web Services\ based back-end processing for visual query generation, transmission and similarity detection demonstrate that the proposed model forms a framework that accurately incorporates the effect of various system parameters with respect to energy consumption and cloud billing costs. Therefore, variations of the proposed analytic modeling can be used for early-stage analysis of possible deployments, or limit studies of the expected performance under a wide range of parameter settings, prior to costly deployments in the field.

\section{Appendix}

\subsection{Proof of Proposition \ref{prop:monotonicity}}

We observe that $E_{\mathrm{exp}}$ is strictly-increasing in $c_{\mathrm{e}}$, since $\frac{d E_{\mathrm{exp}}}{d c_{\mathrm{e}}}  > 0$ for all values of $c_{\mathrm{e}}$ larger than the left extremum of the support of  $\Psi_{\mathrm{e}}$. Moreover, $E_{\mathrm{var}}$ is strictly-decreasing in $c_{\mathrm{e}}$. In order to prove this, we express the dependence of $E_{\mathrm{var}}$ from $c_{\mathrm{e}}$ by using the notation $E_{\mathrm{var}}(c_{\mathrm{e}})$, and we consider two values $c_{\mathrm{e}}'\geq 0$ and $c_{\mathrm{e}}''\geq 0$ such that $c_{\mathrm{e}}' > c_{\mathrm{e}}''$. Then, 
\begin{IEEEeqnarray}{rCl}
E_{\mathrm{var}} (c_{\mathrm{e}}' ) & = & g_{\mathrm{e}}^2 \int_{c_{\mathrm{e}}' E[\Psi_{\mathrm{e}}]}^{+\infty}  (\psi_{\mathrm{e}} - c_{\mathrm{e}}' E[\Psi_{\mathrm{e}}])^2 P_{a}(\psi_{\mathrm{e}}) d\psi_{\mathrm{e}} \\
 & \leq & g_{\mathrm{e}}^2 \int_{c_{\mathrm{e}}'' E[\Psi_{\mathrm{e}}]}^{+\infty}  (\psi_{\mathrm{e}} - c_{\mathrm{e}}' E[\Psi_{\mathrm{e}}])^2 P_{a}(\psi_{\mathrm{e}}) d\psi_{\mathrm{e}} \\
& < &  g_{\mathrm{e}}^2 \int_{c_{\mathrm{e}}'' E[\Psi_{\mathrm{e}}]}^{+\infty}  (\psi_{\mathrm{e}} - c_{\mathrm{e}}'' E[\Psi_{\mathrm{e}}])^2 P_{a}(\psi_{\mathrm{e}}) d\psi_{\mathrm{e}}\\
&= & E_{\mathrm{var}} (c_{\mathrm{e}}'' ),
\end{IEEEeqnarray}
where the first inequality follows from integrating a positive function over a subset, and the second inequality follows from $(\psi_{\mathrm{e}} - c_{\mathrm{e}}' E[\Psi_{\mathrm{e}}])^2 > (\psi_{\mathrm{e}} - c_{\mathrm{e}}'' E[\Psi_{\mathrm{e}}])^2$, when $\psi_{\mathrm{e}} \geq c_{\mathrm{e}}'' E[\Psi_{\mathrm{e}}]$.

The monotonicity properties of $E_{\mathrm{exp}}$ and $E_{\mathrm{var}}$ imply that the constraints in \eqref{eq:primal} and \eqref{eq:dual} are active at the optimum point. Therefore, on recalling that such optimization problems are convex, complementary slackness \cite{Boyd04} implies that the solution of the problem \eqref{eq:primal} is such that $E_{\mathrm{exp}} = E_{\mathrm{max\_exp}}$ and the solution of \eqref{eq:dual} is such that $E_{\mathrm{var}} = E_\mathrm{max\_var}$.

\bibliographystyle{IEEEtran}
\bibliography{literatur}

\end{document}